\documentclass[12pt,preprint]{aastex}

\usepackage{epsfig,amssymb}
\usepackage{graphicx}
\usepackage{rotating}
\usepackage{mathrsfs}
\usepackage{amssymb,amsmath}
\newcommand{\ben}{\begin{enumerate}}
\newcommand{\een}{\end{enumerate}}
\newcommand{\bfig}{\begin{figure}}
\newcommand{\efig}{\end{figure}}
\newcommand{\beq}{\begin{equation}}
\newcommand{\eeq}{\end{equation}}

\newcommand{\mcal}{\mathcal}

\def\degr{\hbox{$^{\circ}$}}

\shorttitle{Magnetic Energy and Helicity Budgets in NOAA AR 11158}

\shortauthors{Tziotziou, Georgoulis, \& Liu}

\begin{document}

\title{Interpreting Eruptive Behavior in NOAA AR 11158 via the Region's Magnetic Energy and Relative Helicity Budgets}

\author{Kostas Tziotziou \& Manolis  K. Georgoulis\altaffilmark{1}}
\affil{Research Center for Astronomy and Applied Mathematics (RCAAM)\\
       Academy of Athens, 4 Soranou Efesiou Street, Athens, GR-11527, Greece}

\and
\author{Yang Liu}
\affil{W.W. Hansen Experimental Physics Laboratory, Stanford University,\\
Stanford, CA 94305-4085, USA}

\altaffiltext{1}{Marie Curie Fellow.}

\begin{abstract}
In previous works we introduced a nonlinear force-free method that
self-consistently calculates the instantaneous budgets of free magnetic
energy and relative magnetic helicity in solar active regions (ARs).
Calculation is expedient and practical, using only a single vector
magnetogram per computation. We apply this method to a timeseries of 600
high-cadence vector magnetograms of the eruptive NOAA AR 11158 acquired by
the Helioseismic and Magnetic Imager onboard the Solar Dynamics Observatory
over a five-day observing interval. Besides testing our method extensively,
we use it to interpret the dynamical evolution in the AR, including
eruptions. We find that the AR builds large budgets of {\it both} free
magnetic energy and relative magnetic helicity, sufficient to power many
more eruptions than the ones it gave within the interval of interest. For
each of these major eruptions, we find eruption-related decreases and
subsequent free-energy and helicity budgets that are consistent with the
observed eruption (flare and coronal-mass-ejection [CME]) sizes. In
addition, we find that (1) evolution in the AR is consistent with the
recently proposed (free) energy -- (relative) helicity diagram of solar ARs,
(2) eruption-related decreases occur {\it before} the flare and the
projected CME-launch times, suggesting that CME progenitors precede flares,
and (3) self-terms of free energy and relative helicity most likely
originate from respective mutual-terms, following a progressive
mutual-to-self conversion pattern that most likely stems from magnetic
reconnection. This results in the non-ideal formation of increasingly
helical pre-eruption structures and instigates further research on the
triggering of solar eruptions with magnetic helicity firmly placed in the
eruption cadre.
\end{abstract}


\section{Introduction}
\label{intro}

A physically meaningful way to quantify the non-potentiality or, in other
words, the eruptive capacity of solar active regions (ARs) is by calculating
these regions' non-potential, ``free'' magnetic energy and, perhaps, the
associated (relative) magnetic helicity. The free magnetic energy is the
excess energy stored in the magnetic configuration of ARs, above the
reference, potential energy corresponding to the AR's magnetic energy in the
absence of electric currents. Magnetic helicity, on the other hand, is the
corresponding degree of twist, writhe, and linkage of the AR's magnetic
field lines \citep[][ and references therein]{berg99}. Adopting the
formalism of the {\em relative} magnetic helicity \citep{berg84}, necessary
in case the studied magnetic configuration permeates a given boundary (such
as the photosphere, for ARs), nonzero (left- or right-handed) total magnetic
helicity implies nonzero free energy. On the other hand, zero magnetic
helicity, as a signed quantity, may imply a nonzero free magnetic energy, in
case equal and opposite amounts of left- and right-handed helicity are
simultaneously present in the configuration. Lack of electric currents
automatically means a zero free magnetic energy and a zero relative
helicity, at the same time meaning inability of an AR to energetically power
solar eruptions, that is, flares and coronal mass ejections (CMEs)
\citep[e.g.][ and references therein]{schr09}.

Contrary to free magnetic energy, a possibly decisive role of magnetic
helicity in solar eruptions is strongly debated. While free magnetic energy
dissipates in non-ideal processes such as magnetic reconnection, helicity is
-- to a very good approximation -- conserved in the course of reconnection.
Indeed, helicity dissipation appears proportional to the inverse square of
the magnetic Reynolds number \citep[e.g.,][]{freebe93, berg99}, and this
makes it insignificant for the solar atmosphere given the very large
value of the latter. If not transferred due to reconnection via existing
magnetic connections to larger scales, therefore (i.e. in remote parts of
the global solar magnetic field), then helicity can be bodily removed from
ARs in the form of CMEs \citep{low94, devo00}. In smaller amounts,
helicity can also escape to the heliosphere via unwinding motions of ``open"
magnetic field lines in case helical, ``closed" field lines transfer their
helicity to them via interchange reconnection \citep[e.g.,][]{pariat09,
raouafi10}.

Outstanding questions that might well involve magnetic helicity and its
evolution in the solar AR corona include (1) the triggering of solar
eruptions and (2) the causal relationship between the two eruption aspects,
namely flares and CMEs (a.k.a. the {\it flare-CME connection}). Numerous
observational and modeling efforts have long suggested that ARs with large
helicity budgets tend to be more eruptive than others \citep[e.g.,][]{can99,
nind04, lab07, nind09, geo09, smyr10, tzio12} with some of these works
actually attributing eruption onset to helicity, via the helical kink
instability \citep[e.g.][]{rulab05, tor05, gib06, kum12}. A ``large"
helicity budget implies a {\it dominant} sign of helicity in ARs. On the
other hand, there have been counterarguments over helicity's necessity for
eruptions in successful eruption models that utilize roughly equal and
opposite amounts of left- and right-handed helicity \citep{phil05, zucc09}.

A reasonable step ahead would be to simultaneously and self-consistently
calculate the budgets of free magnetic energy and relative magnetic helicity
in solar ARs. Existing, widely-applied techniques to this purpose involve
time-integration of the relative-helicity and energy injection rates
obtained via the Poynting theorem for helicity and energy, respectively, on
the photospheric boundary \citep[e.g.,][]{berfie84, kusa02} or evaluation of
the relative-helicity formula in the three-dimensional active-region corona
\citep{berg84, finn85}. Both approaches have nontrivial prerequisites:
helicity- and energy-rate calculation require a photospheric flow-velocity
field that involves significant uncertainties \citep[e.g.,][]{wels07},
together with the field-generating vector potential in the lower boundary
\citep{chae01}. Volume-calculation of helicity and energy, on the other
hand, require an extrapolated three-dimensional magnetic field and its
generating vector potential, including the corresponding current-free field
and vector potential. Furthermore, nonlinear force-free (NLFF) field
extrapolations are model-dependent and subject to uncertainties and
ambiguities \citep{schr06,metc08}. Besides the computational expense stemming from 
the current spatial-resolution capabilities of observed solar
vector magnetograms that render these calculations (particularly the volume
calculation) infeasible for routine use, the additional prerequisites
contribute to our overall inability to credibly monitor the magnetic energy
and helicity budgets in solar ARs.

With these limitations in mind, \cite{geolab07} introduced a technique that
calculates the snapshot (instantaneous) free-magnetic-energy and
relative-magnetic-helicity budgets in solar ARs using {\it only} a single
vector magnetogram for each calculation. The calculation is self-consistent
and strictly adopts the relative-helicity formulation (i.e., a zero relative
helicity for a zero free energy). However, this first effort assumed linear
force-free fields that are known to be unrealistic for solar ARs. Recently,
\citet{geo12a} generalized the method to accommodate NLFF fields for the
first time. Again, the method does not require any extrapolations,
velocities, or vector-potential calculations. \cite{tzio12} applied this
method to 162 vector magnetograms of 42 different ARs to construct the
first free-magnetic-energy -- relative-magnetic-helicity (EH) diagram of
solar ARs. They reported a monotonic relation between free energy and
relative helicity, thereby concluding  that eruptive ARs tend to
accumulate large budgets of {\it both} quantities. They further inferred
well-defined thresholds of free energy and relative helicity for ARs to
enter major (at least M-class) flaring and/or eruptive territory. These
thresholds are $4 \times 10^{31}$ erg and $2 \times 10^{42}$ Mx$^2$,
respectively. Calculation is straightforward and needs only a small fraction
of the computational resources needed for a more conventional calculation of
free energy and helicity budgets.

These developments occurred in excellent timing with the advent of the Solar
Dynamics Observatory (SDO) mission \citep{pes12} featuring the Helioseismic
and Magnetic Imager (HMI) vector magnetograph \citep{sche12}. Besides the
constant-quality, high-spatial-resolution SDO/HMI data, the instrument's
distinguishing feature is its unprecedented cadence. This has led to
detailed, lengthy timeseries of AR vector magnetograms acquired within 12
minutes from each other. Such timeseries are ideal for the NLFF
energy/helicity calculation method of \citet{geo12a} in an effort to (a)
test the method extensively, (b) investigate the statistical robustness of
the EH diagram of \cite{tzio12}, and (c) address the aforementioned
outstanding questions regarding free-energy, helicity, and their combined or
distinct role in solar eruptions.

From the so-far released SDO/HMI data sets, NOAA AR 11158 is undoubtedly the
best-studied subject (see Section \ref{liter} below). The intensely eruptive
AR gave the first X-class flare of solar cycle 24, thus ending the unusually
prolonged lull of solar eruptive activity associated with the latest solar
minimum. The SDO/HMI timeseries covered a 5-day interval of the evolution of
the AR, including the major X2.2 flare and a lengthy series of other
eruptive and combined flares.

This work is structured as follows: the SDO/HMI data on NOAA AR 11158 are
described in Section~\ref{data}. A brief overview of existing studies on the
AR is presented in Section~\ref{liter}. Section~\ref{metho} presents a
summary of the methodology of \citet{geo12a} for the calculation of free
magnetic energy and relative magnetic helicity, as well as one of the most
credible methods for deriving these budgets from time-integration of the
respective injection rates. This latter method is used as reference for
comparison with our instantaneous NLFF budget calculation. Section \ref{res}
provides our results in detail, while Section \ref{conc} summarizes the
study, discusses the ramifications of our results, and presents our
conclusions.

\section{Data description}
\label{data}

NOAA active region (AR) 11158 emerged in the eastern solar hemisphere, near
disc center, on 2011 February 11. During its first disc transit, {\em GOES}
1-8$\;$\AA\ X-ray flux detectors registered 56 C-class, 5 M-class, and the
first X-class (X2.2) flare of solar cycle 24 (01:44 UT on 2011 February 15)
as stemming from this AR. The X-class flare was associated with
conspicuous halo-CME and EUV wave events \citep{schr11}. The AR evolved
quickly from a simple dipole to a complex, predominantly quadrupolar
configuration with an enhanced and strongly sheared polarity inversion line
(PIL). Snapshots of this rapid, dynamical evolution are presented in
Figure~\ref{ar11158}. Extensive descriptions of the AR's configuration and
evolution can be found in  \cite{schr11}, \cite{jiang12}, \cite{liuc12} and
\cite{vema12a}, while \cite{sun12a} provided a detailed description of the
long-term evolution of the AR, its magnetic field, and observed shear
motions along the main PIL. A discussion of the most important works in the
AR is presented in Section \ref{liter}.

For our analysis we use cutout vector magnetic field data taken with the HMI
onboard SDO. HMI is a full disk (4096$\times$4096) filtergraph with a
0.5\arcsec\ pixel size that samples the \ion{Fe}{1} 617.3 nm photospheric
line through a 76 m\AA\ filter at six wavelength positions along the line,
thus covering a range of $\lambda_{\rm 0}\pm$17.5 pm. The first of HMI's two
CCD cameras records filtergrams at these six wavelength positions in two
polarization states and uses them to obtain Dopplergrams and line-of-sight
(LOS) magnetograms at a cadence of 45 s. The second camera acquires six
polarization states at these six wavelength positions every 135 s and uses
them to compute the four Stokes parameters ({\em I, Q, U,} and {\em V})
necessary to derive the vector magnetic field. This is accomplished by a
Milne-Eddington-based inversion approach \citep{borr11}. Averaged 720-second
filtergrams are used for the derivation of the Stokes parameters in order to
enhance the signal-to-noise ratio and suppress 5-min p-mode oscillations.

We use 600 vector magnetograms of NOAA AR 11158 covering a five-day period
(2011 February 12-16) with a 12-minute cadence. The photospheric area
covered by these magnetograms is 650$\times$600 pixels, or
325\arcsec$\times$300\arcsec on the image (observer's) plane.
Figure~\ref{ar11158} provides a sequence of snapshots and a detailed vector
magnetogram of the region. The vector magnetogram data were acquired by
SDO's Joint Science Operations Center (JSOC). Although the released vector
magnetograms were already treated for the azimuthal 180\degr\ ambiguity, for
reasons explained in \cite{georg12} they were re-processed using the
non-potential field calculation (NPFC) of \citet{geo05}, as revised in
\citet{metc06}. For our analysis we use the heliographic components of
the magnetic field vector on the heliographic plane, derived with the
de-projection equations of \cite{gary90}. As typical single-value
uncertainties for the line-of-sight and transverse field components and for
the azimuth angle ($\delta B_l$, $\delta B_{tr}$ and $\delta \phi$
respectively), we use the mean value of the JSOC-provided respective
uncertainties of all pixels with a line-of-sight field $B_l$ lying within
30\% of the magnetogram's maximum/minimum $B_l$ value. This guarantees
the exclusion of numerous quiet-Sun pixels within the field of view, with
$B_l$ values close to the derived errors, that would contribute with
unrealistically high for an AR magnetic-field and azimuth-angle
uncertainties. The mean values of the derived errors are $\delta
B_l\sim32.5$ G, $\delta B_{tr}\sim38.2$ G and $\delta\phi\sim2.9$\degr.

Figure~\ref{flux} shows the evolution of the unsigned (total) magnetic flux
(thick black curve) and the corresponding positive- and negative-polarity
fluxes (red and blue curves, respectively) over the observing interval. The
thick orange curve denotes the unsigned magnetic flux participating in the
magnetic connectivity matrix of the AR (see Section~\ref{newmethod} for a
definition) which is necessary for our calculations. Notice that the AR was
approximately flux-balanced, with the positive-polarity flux showing a small
($\sim$ 5\% on average) excess over its negative-polarity counterpart. The
dynamical evolution of the AR kicks in after decimal day 12.8 ($\sim $19:00
UT on 2011 February 12) and is signaled by the abrupt increase of all flux
budgets, particularly the unsigned connected flux (orange curve). Subsequent
increases and decreases of this flux budget reflect the physical evolution
in the AR, from flux emergence and enhancement of the connectivity matrix to
flaring and eruptions (see Section~\ref{evol} for a detailed description).
During the observing interval the AR released one X-class, 3 M-class and 25
C-class flares.

\section{NOAA AR 11158 in literature}
\label{liter}

NOAA AR 11158 has already been the subject of numerous studies, using a
variety of methods and observations, and also including the same HMI data
set used for the present study. \cite{schr11} presented the first detailed
work, supported by MHD modeling, on several aspects of the X2.2 flare of
February 15, 2011 such as the study of expanding loops above the PIL, the
observed EIT wave event and the accompanying halo CME. The evolution of
magnetic field and magnetic energy in NOAA AR 11158 were discussed in detail
by \cite{sun12a} with the use of NLFF field extrapolations. More recently,
\cite{sun12b} focused on the M2.2 flare triggered on February 14, 2011 and
used NLFF field extrapolations to study non-radial eruptions modulated by
the local magnetic field geometry and the existence of a coronal null point.

Several studies focused on horizontal fields, shear motions along the PIL,
Lorentz-force calculations, and peculiar flows (proper motions) in NOAA AR 11158. A
NLFF field-computation of the M6.6 flare of February 13 \citep{liuc12}
showed an 28\%-increase of the mean horizontal field near
the magnetic polarity inversion line (PIL) and a downward collapse of a
strong horizontal current system above the AR's photosphere after the flare.
\cite{wang12} reported a similar result for the X2.2 flare, showing a rapid,
irreversible enhancement (30\%) of the horizontal magnetic field at the PIL
and an increase of both inclination and shear of the photospheric field as a
response to the flare. This agrees with the results of
\cite{sun12a} who used a NLFF field extrapolation to show an
enhancement of the horizontal field that becomes both more inclined and more
parallel to the PIL. \cite{gosa12} also found that the field became more
horizontal after the flare close to the PIL. \cite{jiang12}
detected a clockwise rotation in the sunspot, 20 hours prior to the X-class
flare that stopped one hour after the flare and suggested that these motions
were associated with the buildup of helicity in the region and the
shearing of the main neutral line. Abrupt changes during the X2.2 flare in
the vertical electric current, the Lorentz force vector and the field
vector, that became stronger and more horizontal, were also described by
\cite{petrie12}; these changes were mainly concentrated near the PIL where
the shear increased. \cite{alva12} used HMI dopplergram data to identify,
through standard methods of local helioseismology, seismic sources in the
X2.2 flare event. They correlated these sources with the HMI magnetic-field changes
and estimated the work done by the Lorentz force
integrated over the entire AR. \cite{liuy12a} studied photospheric flows
derived from HMI vector fields and sub-photospheric flows derived by
time-distance helioseismology from HMI dopplergrams to conclude that
horizontal flows associated with flux emergence, including apparent shear
motion along the PIL, do not extend deeply into the subsurface. Horizontal
proper motions of a few 100 m~s$^{-1}$ along the main neutral line of the
X2.2 flare were measured by \cite{beau12} using a local correlation
tracking method on HMI continuum images and longitudinal magnetograms.
\cite{maur12} further reported magnetic and Doppler transients and locations of
spectral line reversals during the flare's impulsive phase that do not
correspond to real changes of the photospheric magnetic and velocity field.

Concerning helicity budgets in NOAA AR 11158, \cite{vema12a} studied the
helicity injection (through shuffling) and its behavior and found that
changes of the helicity flux signal due to injection of negative helicity in
regions of dominant positive helicity are co-spatial and co-temporal with
flaring/eruptive sites. \cite{jing12}, using the same HMI vector magnetogram
data set as our study, calculated the average current helicity as a function
of altitude and relative helicity with a NLFF coronal field extrapolation
method. \cite{nind12}, in a study of the role of the overlying background
field in solar eruptions, also calculated from the same HMI vector
magnetogram data the helicity flux and budget in NOAA AR 11158. Electric
current, fractional current helicity (i.e., helicity due to vertical
electric currents), photospheric free energy, and angular shear were also
recently derived by \cite{song12} in an attempt to quantify the
non-potentiality of NOAA AR 11158. \cite{vema12b}, on the other hand,
studied two sub-regions of the AR with rotating sunspots and showed that
proper and rotational motions play a key role in enhancing the magnetic
non-potentiality of the AR by injecting helicity and twisting the magnetic
fields, thus increasing the free energy.

Recently, \cite{liuy12b}, in a comprehensive study of magnetic helicity and
energy in NOAA AR 11158, studied the region's evolution
by deriving and integrating
helicity injection from both shuffling and emergence.
They concluded that helicity is
mainly contributed by shuffling while the energy build-up is mainly
attributed to emergence.

Finally, a study of loops in NOAA AR 11158 was carried out by \cite{asch11},
who applied an automated analysis of SDO/AIA image datasets, to detect 570
loop segments, shortly before the X-class flare, at temperatures covering
the range of 10$^{5.7}$ - 10$^{7}$ K. \cite{gosa12} used SDO/AIA
observations to also study the evolution of coronal loops during the X2.2
flare which show three distinct dynamical phases; a slow rise phase prior to
the flare, a collapse phase during the flare and an oscillation phase
following the flare-driven implosion.

\section{Methodology: free magnetic energy and relative magnetic helicity budgets}
\label{metho}

In this Section we present a) our method for calculating the instantaneous
free magnetic energy and relative magnetic helicity budgets in an AR
(Section~\ref{newmethod}), and b) the derivation of these budgets by
integrating the respective energy/helicity fluxes inferred via the
photospheric velocity field (Section~\ref{dave4vm}). We emphasize that
the main analysis method is the one described in Section \ref{newmethod};
integrated helicity and energy fluxes are used as a reference for comparison
with our results.

\subsection{NLFF energy and helicity budgets from single vector magnetograms}
\label{newmethod}

Recently, \citet{geo12a} proposed a new NLFF method to derive the
(instantaneous) free magnetic-energy and relative magnetic-helicity budgets
using a single photospheric or chromospheric vector magnetogram of the
studied AR. This method provides unique results, contrary to model-dependent
NLFF field extrapolation methods. In return, it requires a unique
magnetic-connectivity matrix, i.e., a matrix containing the flux committed
to connections between positive- and negative-polarity flux partitions. This
matrix is calculated using a simulated annealing method
\citep{geo:rus,geo12a} that guarantees connections between opposite-polarity
flux partitions while globally minimizing the corresponding connection
lengths. The connectivity matrix defines a collection of $N$ magnetic
connections, treated as slender force-free flux tubes with known footpoints,
flux contents, and variable force-free parameters.

For this collection of flux tubes, the free magnetic energy $E_c$ is the sum
of a self term $E_{c_{\rm self}}$, expressing the internal twist and writhe
of each tube, and a mutual term $E_{c_{\rm mut}}$, due to interactions
between different flux tubes. As such, it is given by
\begin{eqnarray}
E_c  & = & E_{c_{\rm self}} + E_{c_{\rm mut}} \nonumber \\
 & = & A d^2 \sum _{l=1}^N \alpha _l^2
\Phi_l^{2 \delta} +
      \frac{1}{8 \pi} \sum _{l=1}^N \sum _{m=1, l \ne m}^N
           \alpha _l \mcal{L}_{lm}^{\rm arch} \Phi_l \Phi_m\;\;,
\label{Ec_fin}
\end{eqnarray}
where $A$ and $\delta$ are known fitting constants, $d$ is the pixel size of
the magnetogram and $\Phi_l$ and $\alpha_l$ are the respective flux and
force-free parameters of flux tube $l$. $\mcal{L}_{lm}^{\rm arch}$ is the
mutual-helicity factor of two arch-like flux tubes that do not wind around
each other's axes. Inference of this factor, first introduced by
\citet{dem06}, is discussed in detail by \cite{geo12a} for all possible
cases of intersecting/non-intersecting flux tubes pairs and flux-tube pairs
with a ``matching" footpoint, per the reduced spatial resolution of the
flux-partitioning map. Notice that the free energy of Equation
(\ref{Ec_fin}) is to be taken as a {\em lower limit} of the actual energy as
the unknown winding factor around flux tubes, that would require
knowledge of the three-dimensional magnetic field, has been set to zero.
Previous works \citep{geo12a, tzio12} have shown that this lower limit is
realistic.

Likewise, the respective relative magnetic helicity $H_m$ is the sum of a
self $H_{m_{\rm self}}$ and a mutual $H_{m_{\rm mut}}$ term,
\begin{eqnarray}
H_m  & = & H_{m_{\rm self}} + H_{m_{\rm mut}} \nonumber \\
&  = & 8 \pi d^2 A
\sum _{l=1}^N \alpha _l \Phi_l ^{2 \delta} +
      \sum _{l=1}^N \sum _{m=1,l \ne m}^N \mcal{L}_{lm}^{\rm arch} \Phi_l
      \Phi_m\;\;.
\label{Hm_fin}
\end{eqnarray}
We refer the reader to \citet{geo12a} for a detailed description of the
method and the derivation of uncertainties for all terms of Equations
(\ref{Ec_fin}) and (\ref{Hm_fin}).

\subsection{NLFF energy and helicity budgets by integration of respective fluxes}
\label{dave4vm}

A method to infer the (total) energy and relative helicity budgets in an
active region at time $t$ is by integrating the respective fluxes from
initiation $t_{\rm 0}$ of the active region to time $t$. The energy and
helicity budgets in this case are given by
\begin{equation}
E \equiv E(t) = \int _{t_0}^t (dE/dt) dt \label{enfl}
\end{equation}
and \begin{equation} H_m \equiv H_m(t) = \int _{t_0}^t (dH_m/dt) dt,
\label{hmfl}
\end{equation}
respectively, where $E(t_{\rm 0})=H_m(t_{\rm 0})=0$. The helicity injection
rate \citep{berfie84} depends on a) the photospheric velocity vector, whose
inference involves significant uncertainties \citep[e.g.,][]{wels07}, and b)
the vector potential $\mathbf{A}_p$ of the potential magnetic field
$\mathbf{B}_p$ exhibiting the same normal-field condition as the observed
field  $\mathbf{B}$.

In the simplest case, photospheric velocities can be inferred from a
sequence of magnetograms through local correlation tracking
\citep[LCT;][]{nov88} techniques which, however, are highly uncertain due to
the ambiguities in image motion. Moreover, LCT techniques tend to
underestimate the amount of helicity injected by the shear term
\citep{dem03} and are inconsistent with the magnetic induction equation,
which governs the evolution of the photospheric magnetic fields
\citep{schuck05}.

Several alternative methods have been proposed during the past decade
\citep[see ][and references therein]{wels07} for the evaluation of plasma
velocities. Here we use the
Differential Affine Velocity Estimator for Vector Magnetograms
(DAVE4VM) introduced by \citet{schuck08}, which is a generalization of the
Differential Affine Velocity Estimator \citep[DAVE;][]{schuck05,schuck06}
method. DAVE4VM estimates the plasma velocity $\boldsymbol{\upsilon}$ using
the normal component of the ideal induction equation
\begin{equation}
\partial_t{B_z} + \nabla_h \cdot (B_z \boldsymbol{\upsilon}_h - \upsilon_z\mathbf{B}_h) = 0\;\;,
\end{equation}
where $\mathbf{B}$ is the magnetic field vector, $z$ denotes the normal
(vertical) axis and $h$ denotes $x$- and $y$-components on the tangential
(horizontal) plane. We apply this method to our timeseries of vector
magnetograms on the heliographic plane, using a window size of 21 pixels in
DAVE4VM. Since the coordinate system for the affine velocity profile is not
aligned with the magnetic field (velocities not necessarily orthogonal to
$\mathbf{B}$), the calculated velocity $\boldsymbol{\upsilon}$ has to be
further corrected by removing the field-aligned plasma flow
$\boldsymbol{\upsilon}_\parallel$ and keeping only the perpendicular
($\perp$) component that plays a role in the induction equation. The
corrected cross-field velocity is given by
\begin{equation}
\boldsymbol{\upsilon}_\perp = \boldsymbol{\upsilon} -
\frac{(\boldsymbol{\upsilon} \cdot \mathbf{B})\mathbf{B}}{B^2}\;\;.
\end{equation}
The normal $\upsilon_{\perp n}$ and tangential $\upsilon_{\perp t}$
components of this velocity $\boldsymbol{\upsilon}_\perp$ are then used to
calculate the helicity flux $\frac{dH}{dt}\Big|_S$ across the plane $S$ of
the magnetogram. As \citet{berfie84} have shown, this is derived by
\begin{equation}
\frac{dH}{dt}\Bigg|_S = 2 \int_S ( \mathbf{A}_p \cdot \mathbf{B}_h )
\upsilon_{\perp n} dS - 2 \int_S ( \mathbf{A}_p \cdot
\boldsymbol{\upsilon}_{\perp t} ) B_z dS\;\;, \label{hmdv}
\end{equation}
where the first term roughly corresponds to the helicity flux through
emergence and the second term to helicity flux through photospheric
shuffling. The vector potential $\mathbf{A}_p$ is calculated by means of a
fast Fourier transform method, as implemented by \citet{chae01}. We refer
the reader to \cite{liuy12b} for a discussion concerning issues related to
the computation and interpretation of
helicity fluxes related to tangential and vertical flows.

Likewise, the magnetic-energy injection rate (i.e., the Poynting flux)
$\frac{dE}{dt}\Big|_S$, as \cite{kusa02} have shown, is given by
\begin{equation}
\frac{dE}{dt}\Bigg|_S = \frac{1}{4\pi}  \int_S B^{2}_h \upsilon_{\perp n} dS
- \frac{1}{4\pi} \int_S ( \mathbf{B}_h \cdot \boldsymbol{\upsilon}_{\perp t}
) B_z dS\;\;, \label{ecdv}
\end{equation}
where the first term roughly corresponds to the energy flux through
emergence and the second term to energy flux through shuffling.

\section{Results}
\label{res}

\subsection{Helicity and energy budgets in NOAA AR 11158}
\label{ehbudget}

We first apply the method described in Section \ref{newmethod} to the 600
SDO/HMI magnetograms of our sample. Figures~\ref{hmevol} and \ref{ecevol}
show respectively the relative magnetic helicity and the free magnetic
energy budgets in the AR as functions of time. Totals (panels [b]) and
self-terms (panels [c]) are also shown. Given the small amplitudes of the
self terms for energy and helicity, the corresponding mutual terms are
indistinguishable from the totals and are hence not shown here.
Figure~\ref{ecevol} also shows the potential and total magnetic energy and
their evolution. Uncertainties are shown by grey bars in all plots. For the
total and mutual energy/helicity terms these uncertainties are relatively
small, while they are relatively large for the self terms.

All relative-helicity and free-energy terms show a relatively smooth,
progressive evolution, with some scatter around the corresponding
72-min-running averages. Scatter increases after decimal day 13.5 when the
AR starts building significant budgets of magnetic helicity and free energy
powered by continuous flux emergence (Figure~\ref{flux}). An important
finding is that the region builds {\em both} significant relative
magnetic helicity and free magnetic energy, reaching peaks of $\sim 2
\times 10^{43}$ Mx$^2$ and $\sim 5.5 \times 10^{32}$ erg respectively. These
budgets are large enough to power several eruptive flares which, in fact, is
what the AR did over the 5-day observation interval (Section \ref{data}).
The peak times of all these flares are marked in Figures~\ref{hmevol},
\ref{ecevol}. Eruptive flares can be assessed fairly easily by using
frequency-time radio spectra such as those of Figures~\ref{hmevol}a,
\ref{ecevol}a, obtained by the WAVES instrument onboard the Wind mission
\citep{boug95}. Type-II radio activity in particular, corresponding to
hours-long frequency drift of the radio emission, is a tell-tale signature
of shock-fronted CMEs  \citep[see, e.g.,][ and references therein]{hill12}.
Much faster (shorter) frequency drifts (Type-III bursts) typify
magnetic-reconnection episodes in the solar atmosphere. Association of
registered GOES flares with observed CMEs is based on a) correspondence of
the peak time of the flare with the onset time of the Type-II radio burst,
and b) available CME catalogues, such as the LASCO, CACTUS and SEEDS. This
is because neither {\em GOES} nor {\em Wind}/WAVES have spatial resolution.
While the flare-CME connection is discussed in Section \ref{erupt} in more
detail, from Figures ~\ref{hmevol}a, \ref{ecevol}a we discern that at
least the M6.6 flare of February 13, the M2.2 flare of February 14, the X2.2
flare of February 15, and the M1.6 flare of February 16 were eruptive.

NOAA AR 11158, an intensely flaring/eruptive AR shows both a significant
budget and a dominant sense of magnetic helicity, positive (right-handed) in
this case, as already suggested by \citet{tzio12} for numerous ARs. Notably,
as well, no major flare ($\ge$ M1.0) occurs before thresholds of $\sim
4 \times 10^{31}$ erg and $\sim 2 \times 10^{42}$ Mx$^2$ in free magnetic
energy and relative magnetic helicity, respectively, are exceeded, as
concluded by \citet{tzio12}. This point is further discussed in
Section~\ref{EHdiagram}.

Comparison of our results (Figure~\ref{ecevol}) with the respective results
by \cite{sun12a} shows that although the corresponding potential, total and
free energy evolution share some qualitative resemblance before the
occurrence of the X-class flare, there are notable differences in their
evolution afterwards: the magnetic energy derived by \cite{sun12a}
continuously decreases after the flare while in our case free energy
increases for almost a day thereafter. Furthermore, there are quantitative
differences within the entire observing interval with the maximum free
magnetic energy derived by our NLFF method being almost twice that derived
by the NLFF field extrapolation (notice that our free energy is a lower
limit[!]). Similar differences also exist between our magnetic free energy
and the respective results by \cite{nind12}, derived from NLFF field
extrapolations of 16 {\em Hinode} SOT/SP vector magnetograms, that do not
differ substantially from the results by \cite{sun12a}. Such discrepancies
are already noted and discussed by \cite{geo12a}. Finally, the relative
helicity evolution derived by \cite{jing12} using a NLFF field extrapolation
differs from our result both quantitatively and quantitatively, as it is
$\sim$ 50\% lower.

\subsection{Comparison with helicity/energy budgets derived from integration of helicity/energy fluxes}
\label{ehcomp}

\begin{figure}
\includegraphics[width=\linewidth]{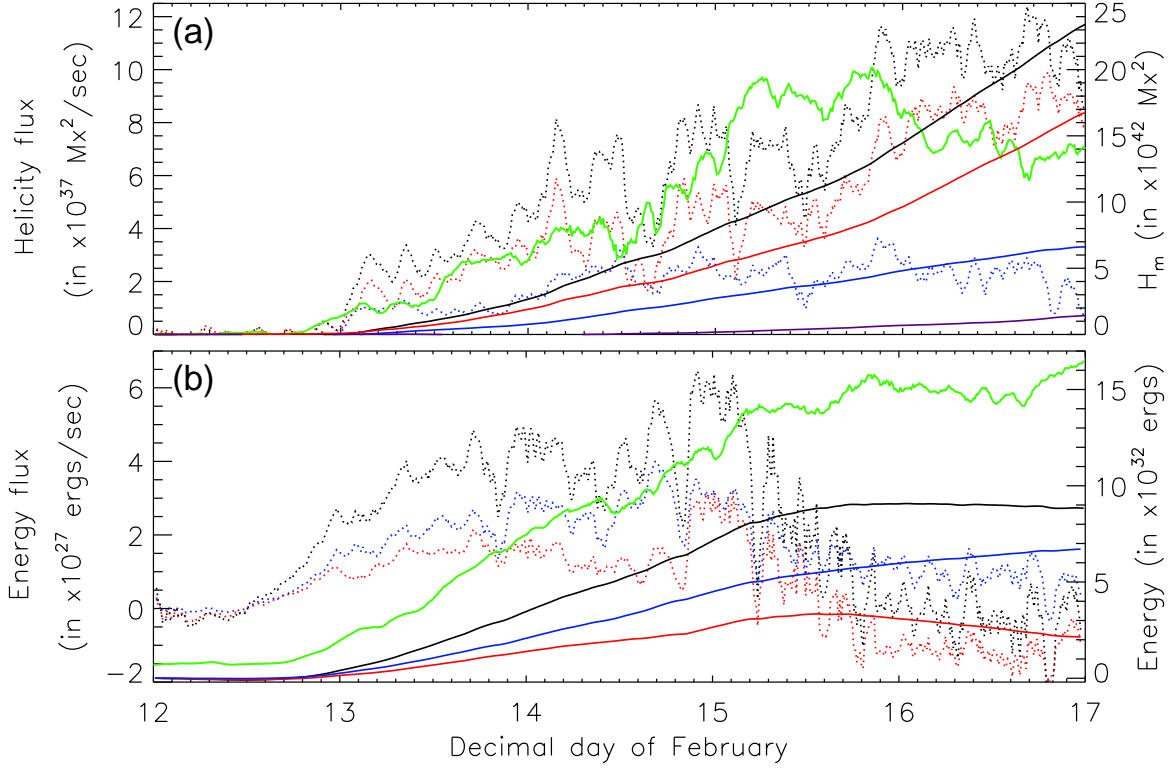}
\caption{72-minute average temporal profiles of the
relative-magnetic-helicity (a) and the total-magnetic-energy (b) fluxes and
their integrated budgets. Dotted red, blue, and black lines correspond to
the shuffling-, emergence-, and total-flux terms, respectively (readings on
left ordinate). The respective integrated budgets as a function of time are
given by solid red, blue, and black lines (readings on right ordinate). For
reference, our NLFF relative helicity budget (a) and total energy budget (b)
are also given by the green curves (readings on right ordinate). The thick
purple line in (a) shows the accumulated emergence helicity inferred using a
LCT velocity (see text).} \label{hmdave}
\end{figure}

Figure~\ref{hmdave} shows the emergence and shuffling terms of helicity
(Figure~\ref{hmdave}a) and energy (Figure~\ref{hmdave}b) fluxes, as well as
the their totals, derived by Equations~(\ref{enfl}) and (\ref{hmfl}).
Emergence and shuffling terms of energy/helicity are always in phase, while
changes in the evolution of both beyond the influence of flux emergence
mainly reflect shear motions along the PIL, separation motions of connected
polarities, and sunspot rotation \citep{liuy12b}. Shuffling dominates over
emergence in helicity injection rates, while emergence dominates over
shuffling in energy injection rates. Moreover, significant changes in the
energy flux seem to occur immediately after the X-class flare (early on
day 15) while there seems to be a $\sim$ 24-hour delay in the helicity flux
response. As discussed in Section~\ref{dave4vm}, magnetic-energy and
helicity budgets at a given time are derived here by integrating the
respective fluxes from the start of observation to the time of interest.
These budgets should be good approximations of the actual budgets because
the start of the observing interval corresponds practically to the birth
($t_{\rm 0}$) of the AR.

Integrating the helicity injection rates (Figure~\ref{hmdave}a) we find a
total accumulated helicity $\sim 2.4 \times 10^{43}$ Mx$^2$, with almost two
thirds of it due to shuffling. This is in agreement with the results
of \cite{liuy12b}. In terms of energy, we find a total accumulated value
$\sim 9 \times 10^{32}$ erg, with $\sim$ 77\% of it due to emergence;
this is slightly different than the percentages reported by \cite{liuy12b}.
We also note that both the shuffling helicity flux profile and the
corresponding accumulated helicity are in fairly good agreement with the
corresponding calculations of \cite{vema12a} who used DAVE instead of
DAVE4VM to derive the velocity flow field in the AR. However, \cite{nind12},
who also used DAVE for deriving the helicity flux, find an accumulated
helicity which is almost 50\% lower than our accumulated helicity due to
shuffling.

The helicity and energy flux curves derived by \cite{liuy12b} show a
qualitative agreement but significant quantitative differences with our
helicity/energy-flux results. Differences stem mostly from a difference in
the field-of-view (FOV) since these authors used a significantly
smaller FOV that tightly encloses the AR. This difference in the FOV
also explains the difference between our unsigned flux (Figure~\ref{flux})
and the corresponding flux derived by \cite{liuy12b} and \cite{sun12a} who
used the same (albeit Lambert de-projected) magnetograms. Use of a $\sim$
200-300 G threshold in our flux derivation, which is equivalent to
restricting the FOV very close to the AR, results in a curve that is
identical to theirs (not shown here). Further small discrepancies in the
helicity flux could result from a) uncertainties in the velocity
calculations, b) differences in the azimuth disambiguation methods,
and c) differences in the de-projection of the magnetograms; as already
mentioned, \cite{liuy12b} used the Lambert (cylindrical equal area)
de-projection method while we use the heliographic de-projection
\citep{gary90}.

Comparing the flux-integrated energy and helicity budgets from Section
\ref{dave4vm} with our respective NLFF budgets from Section \ref{newmethod}
we find a fairly close agreement between the peaks, particularly for the
relative helicity ($\sim 2 \times 10^{43}$ Mx$^2$ [instantaneous] vs. $\sim
2.4 \times 10^{43}$ Mx$^2$ [flux-integrated]) and less so for the total
magnetic energy ($\sim 16 \times 10^{32}$ erg [instantaneous] vs. $\sim 9
\times 10^{32}$ erg [flux-integrated]). Comparing the respective temporal
profiles is not straightforward, however, and could even be misleading: in
our {\em instantaneous} NLFF budgets, activity-related changes reflected on
the connectivity matrix incur a direct impact on the budget values.
Flux-integrated budgets, on the other hand, will not show an
activity-related change unless the photospheric velocity field, primarily,
is impacted by this change. This explains the smoothness of the
flux-integrated budgets in Figure \ref{hmdave} that is contrasted to the
numerous transient variations of the (72-minute averaged) instantaneous
budgets. Nonetheless, a point-to-point comparison between the 
integrated helicity/energy fluxes and the respective NLFF budgets yields
significant correlations, at least up to the end of February 15, just before
our instantaneous budgets started decreasing. We find linear (Pearson) and
rank-order (Spearman) coefficients of the other 0.73 and 0.79, respectively,
for relative magnetic helicity, and 0.83 and 0.84, respectively, for total
magnetic energy.

Concluding, we note in passing that helicity- and energy-flux calculations
and associated budgets ``inherit'' the uncertainties of the inferred
photospheric velocity field, including caveats or shortcomings in its
calculation. We repeated the analysis of Section \ref{dave4vm} using a
classical LCT method applied to successive sub-sequences of five consecutive
$B_z$-images. We found that the LCT-calculated emergence helicity budget is
roughly an order of magnitude smaller than that obtained using
DAVE4VM-inferred velocities (Figure~\ref{hmdave}a).

\subsection{Helicity and energy evolution in NOAA AR 11158}
\label{evol}

\subsubsection{Comparison of helicity/energy budgets with unsigned connected flux}

As already discussed in Section~\ref{data} and elsewhere, NOAA AR 11158
exhibited a very rapid and dynamical evolution, with continuous flux
emergence (Figures~\ref{ar11158} and \ref{flux}) and a rapidly evolving
connected flux (Figure~\ref{flux}). Connected-flux changes reflect changes
in the connectivity matrix that arise from flux emergence, flux cancellation
in small-scale reconnection events and/or large-scale flaring/erupting
activity. Figures~\ref{hmevol} and \ref{ecevol} suggest a good agreement
between the unsigned connected flux and the AR's relative helicity/free
magnetic energy budgets with significant linear and rank-order
correlation coefficients, equal to 0.93 and 0.89 for helicity and 0.88 (both
correlation coefficients) for free magnetic energy. It is, however, the
potential energy that shows a remarkable correlation with the unsigned
connected flux with both correlation coefficients $\sim$ 0.99. This suggests
that the unsigned flux evolution correlates well with the global
magnetic-field evolution in the AR while small increases/decreases indicate
local and/or larger-scale changes in the non-potentiality of the magnetic
field configuration resulting, at times, in significant changes in the
relative helicity and/or free magnetic energy budget. Point taken, we will
now further discuss the dynamical evolution of NOAA AR 11158 by examining
the evolution of magnetic helicity and free magnetic energy in more detail.

\subsubsection{Helicity evolution in NOAA AR 11158}
\label{hel-evol}

\begin{figure*}
\centering
\includegraphics[width=\linewidth]{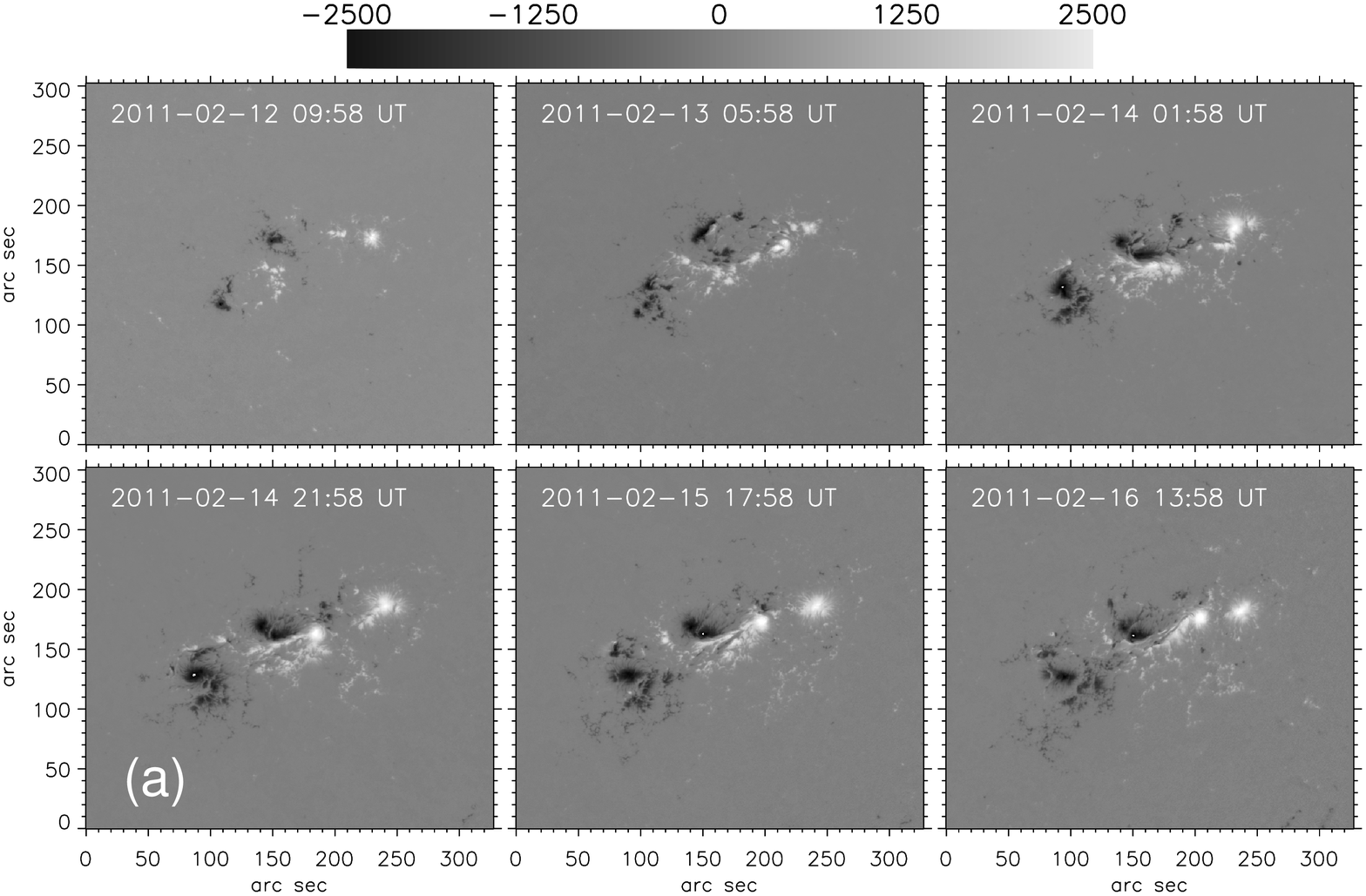}\\
\includegraphics[width=0.8\linewidth]{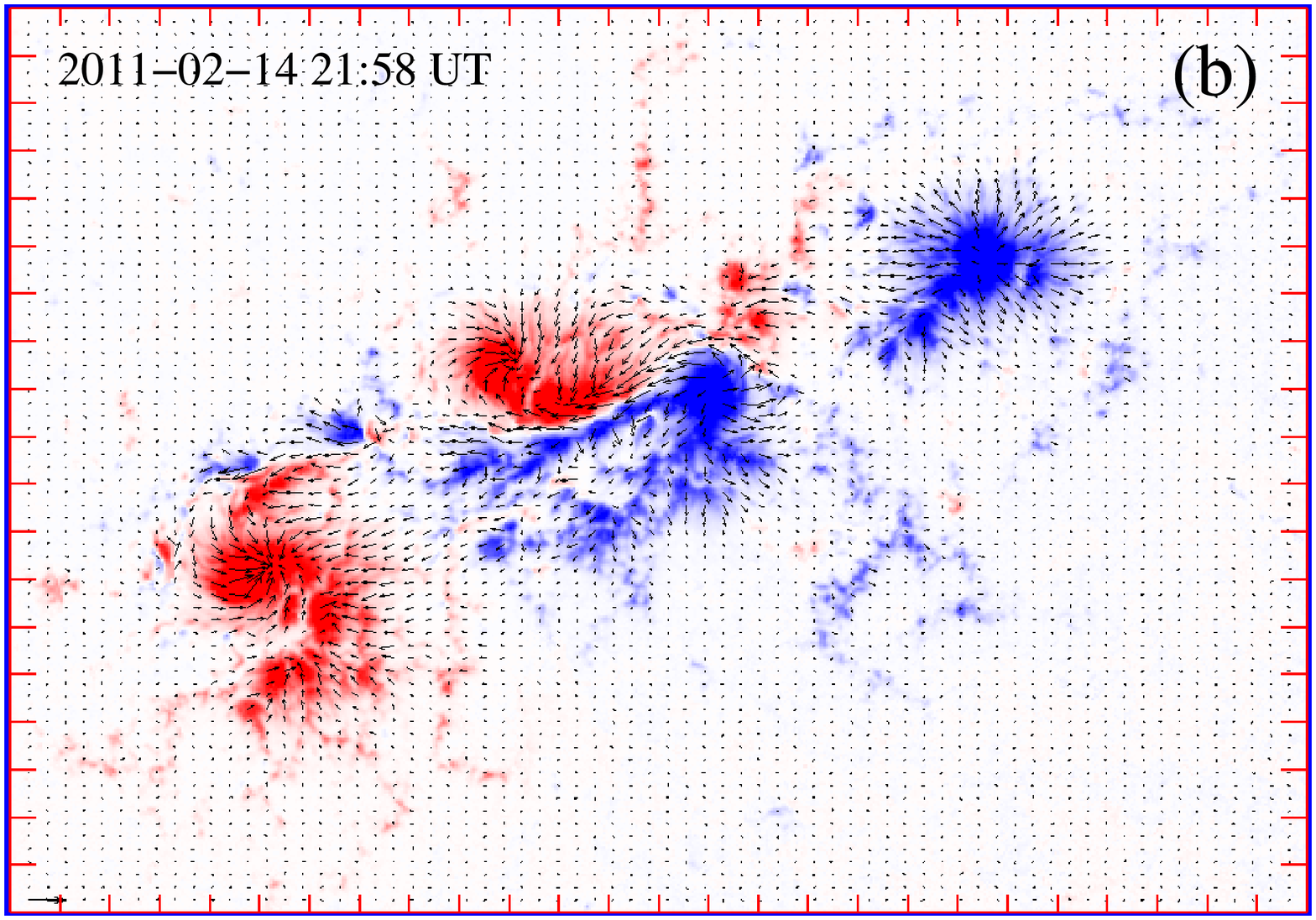}
\caption{(a) Sequence of the heliographic vertical field component
($B_z$) in NOAA AR 11158, as acquired by SDO/HMI. (b) Detailed vector
magnetogram corresponding to the fourth snapshot of the sequence, with $B_z$
saturated at $\pm 1500$ G. The vector in the lower-left edge corresponds to
a horizontal magnetic field of 2000 G. For all images, north is up and west
is to the right.} \label{ar11158}
\end{figure*}

Figure~\ref{hmevol} shows the evolution of relative magnetic helicity in
NOAA AR 11158. In the pre-eruption phase of the AR, particularly during the
first day (February 12, 2011), the relative helicity budget is very low. It
reflects the initially simple $\beta$-configuration (Figure~\ref{ar11158})
of the AR, with a magnetic-field configuration apparently close to a
potential configuration. However, the AR evolved rapidly, within a couple of
days, into a more complex $\beta \gamma \delta$-configuration
(Figure~\ref{ar11158}). Consequently, the helicity budget increased
drastically within less than a day starting around the end of February 12, a
period that coincides with the first cluster of two C-class flares and an
eruptive M6.6 flare (see Figures~\ref{hmevol}, \ref{ecevol} and \ref{cmes}).
Several hours after the M-class flare, late on day 13, a significant
decrease of helicity occurs, co-temporarily with a small increase of the
connected flux, which is not associated with any flaring/eruptive
activity. This helicity decrease seems also to be associated with a delayed
by 2 hours increase of the free magnetic energy  (Figure~\ref{ecevol}) 
and probably suggests a large-scale re-organization
of the AR. Thereafter, during the early hours of day 14, coinciding with a
second cluster of four non-eruptive C-class flares (among them a C8.3 and a
C6.6), helicity increases, but this is followed by a very
significant, simultaneous decrease of helicity, free energy and
connected magnetic flux shortly before midday ($\sim$14.4).
Figure~\ref{openflx1} indicates an intense restructuring of the
magnetic-field configuration accompanied by an increase of the fragmentation
in the AR. The former is reflected on the decrease of the connected flux
$\Phi_{\rm conn}$ while the latter is judged by the abrupt increase of the
flux-tube number. ``Open" flux is further increased, implying that
previously connected flux within the AR now closes beyond the AR's limits.

Helicity increases again, during the last half of day 14, when a swarm of
six C-class flares (the largest ones being C9.4, C7 and an eruptive
C6.6 [Figure ~\ref{cmes}]) and an eruptive M2.2 flare occur. An overall
increasing helicity trend, starting around decimal day 14.5, 
interrupted by short, simultaneous decreases of helicity, free energy and
connected flux associated with eruptions (see further discussion in
Section~\ref{cmes}), could well be related to the reported rotation of the
main sunspots \citep{jiang12, vema12a}. Such a rotation twists and/or shears
nearby magnetic field lines, adding further helicity and free/total energy
to the region. Recently, \cite{vema12b} showed clear correspondence, in a
sub-region of NOAA AR 11158, between the rotational profile of the sunspot
and the respective shear and helicity flux. This latest cluster of flares is
followed in the early hours of day 15 by the large eruptive X2.2 flare, the
first X-class flare of solar cycle 24. The peak of the helicity budget
($\sim 2 \times 10^{43}$ Mx$^2$) is first reached three hours after the
X-class flare. The persistence of sunspot rotation after the X-class flare
\citep{jiang12} could explain this further increase of helicity.

It is worth-discussing further that during the aforementioned 1.5-day
interval before the first helicity peak, despite the occurrence of eruptive
flares that removed helicity from the AR, helicity kept on increasing. This
behavior seems to be the end result of a continuous, uneven competition
between new helicity injection in the region, as Figure~\ref{hmdave} also
suggests, and helicity removal through repetitive eruptive activity with the
former overpowering the latter. There are, as expected, occasional decreases
of helicity during this interval associated with eruptive flares, which will
be later discussed in Section~\ref{erupt}, but the magnetograms
(Figure~\ref{ar11158}) attest to a very rapid and intense dynamical
evolution of the AR, with continuous emergence of new magnetic flux. As a
result, both the unsigned flux and the unsigned connected flux keep
increasing throughout the interval of study (Figure~\ref{flux}). The latter
generally follows the aforementioned helicity evolution, increasing rapidly
between decimal days 12.7 to 15 and then fluctuating around a mean value of
$2.4 \times 10^{22}$ Mx.

\begin{figure}
\centering
\includegraphics[width=0.7\linewidth]{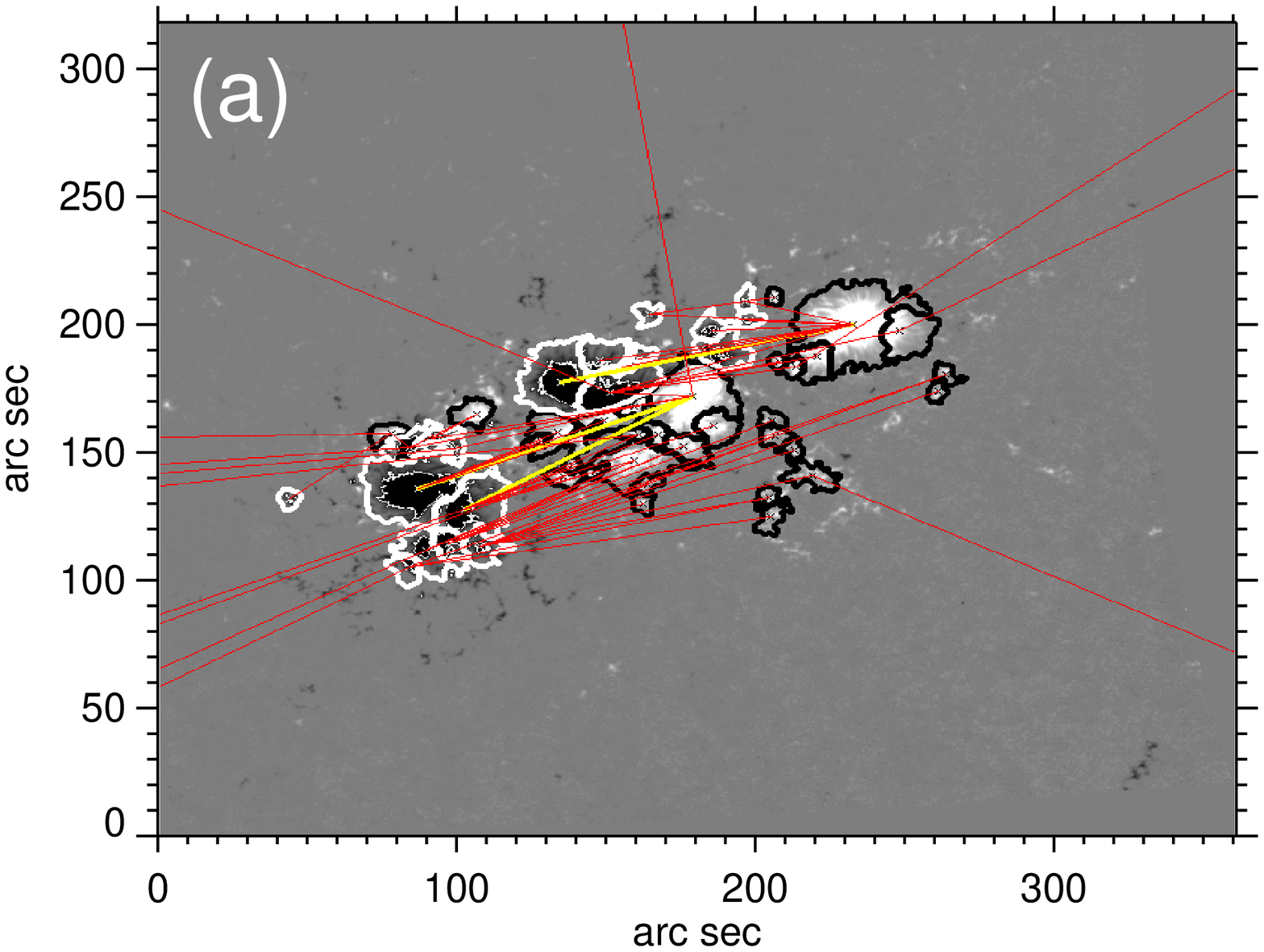}\\
\includegraphics[width=0.7\linewidth]{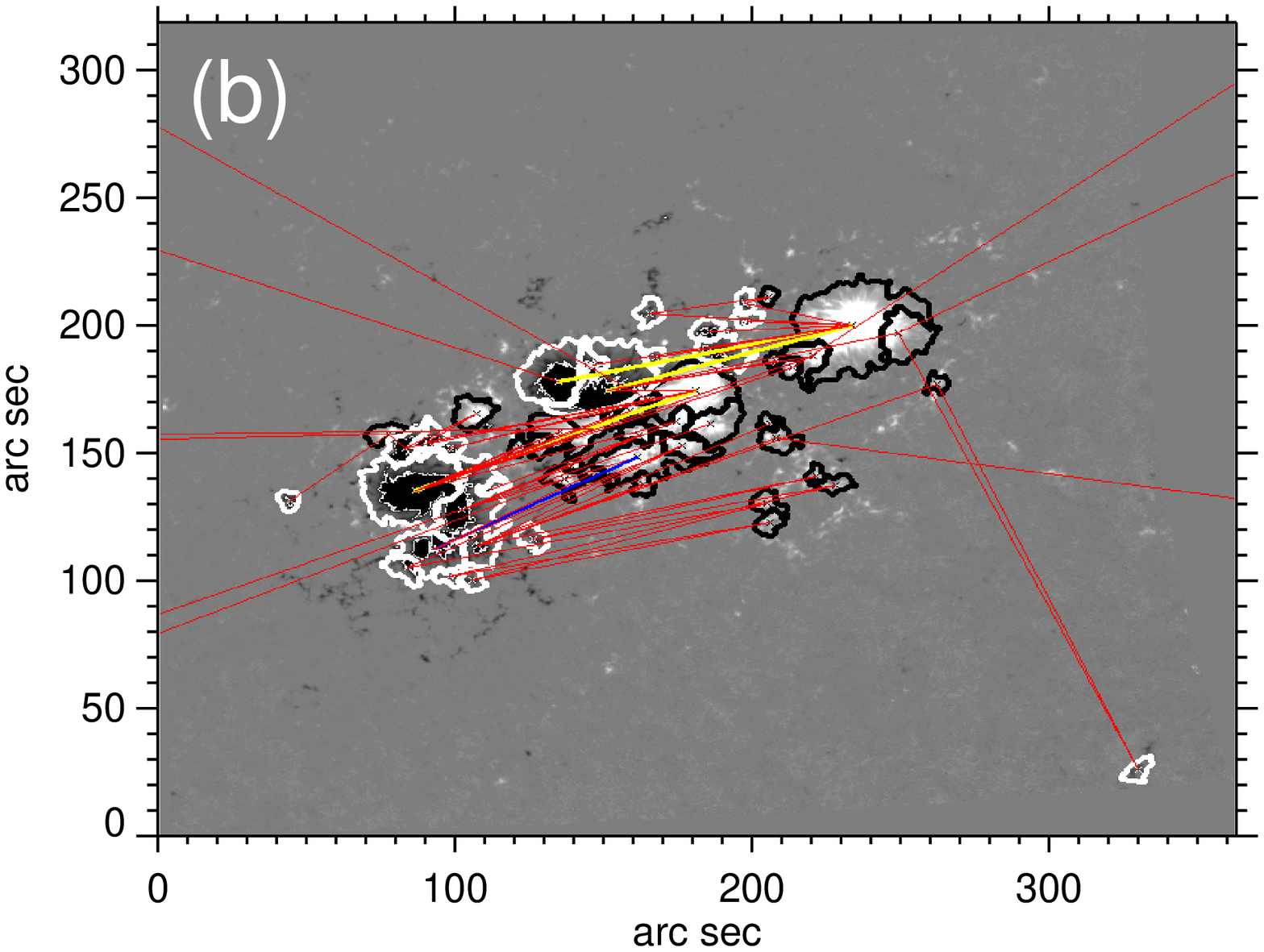}
\caption{Magnetic connectivity in NOAA AR 10254, observed on February 15,
01:24 UT (a) before the onset of the X2.2 flare and on February 15, 02:24 UT
(b) after the flare. Images show the vertical magnetic field component in
gray scale with the contours bounding the identified magnetic partitions.
The flux-tube connections identified by the magnetic connectivity matrix are
represented by line segments connecting the flux-weighted centroids of the
respective pair of partitions. Red, cyan, and yellow segments denote
magnetic flux contents within the ranges [5$\times10^{19}$,
5$\times10^{20}$] Mx, [5$\times10^{20}$, 10$^{21}$] Mx, and [10$^{21}$,
5$\times10^{21}$] Mx, respectively. Both closed and ``open", i.e. closing
beyond the AR's limits, connections are shown. North is up; west is to the
right.} \label{conx}
\end{figure}

After its first peak, the helicity budget shows a considerable decrease that
immediately follows the termination of the sunspot rotation reported by
\cite{jiang12}, and probably indicates another re-organization of the AR's
magnetic connectivity, after a large sequence of flaring/eruptive activity.
This re-organization also relates to the decrease of the unsigned connected
flux and the simultaneous increase of free magnetic energy.
Figure~\ref{conx}, provides a glimpse of this re-organization with two
connectivity matrices before and after the X-class flare, separated by an
hour, that clearly show connectivity changes mainly between the AR's eastern
sunspot complex, where the flare occurred, and the AR's central complex of
sunspots. We also remind the reader that an EIT wave event \citep{schr11}
accompanied the X-class flare, which could affect the AR and its vicinity.
Late on February 15, the helicity budget increases again through two
non-eruptive C-class flares and reaches a second, similar peak at decimal
day 15.84, at which time a C6.6 non-eruptive flare occurs. Thereafter, a
substantial, 30\% helicity decrease occurs within 8 hours, accompanied only
by two small non-eruptive C-class flares. This remarkable helicity decrease
also coincides with a considerable decrease of free magnetic energy and, as
Figure~\ref{openflx3} shows,  a decrease of the connected flux. At the
same time the number of flux tubes that comprise the connectivity matrices
increases, indicating a rapid fragmentation of the AR (more magnetic
partitions, defining a larger number of flux tubes but with less total
connected flux), while the open flux also increases indicating a large-scale
re-organization of the magnetic configuration. It is the cumulative effect
of both processes that leads to these observed helicity and energy decreases
while no eruptive activity takes place.

During the last 17 hours of February 16, 2011 seven non-eruptive C-class
flares (among them a C9.9 and a C7.7) and an eruptive M1.6 flare occur while
the helicity fluctuates around a value of $\sim 1.5 \times 10^{43}$ Mx$^2$.
This well-defined mean helicity, despite a swarm of (non-eruptive) flares,
relates to the absence of major eruptions during this interval - only the
eruptive M1.6 flare shows a clear associated helicity decrease (see
Section~\ref{erupt} and Figure~\ref{cmes}).

\subsubsection{Free magnetic energy evolution in NOAA AR 11158}

Figure~\ref{ecevol} shows the evolution of the free magnetic energy budget
in NOAA AR 11158. Free energy generally follows the relative-helicity
evolution, but with some notable differences. During the first four days of
observations, the free magnetic energy shows a similar increase as helicity
and finally peaks less than a day after the X-class flare, at the same time
(decimal day 15.84). The peak energy is $\sim 6 \times 10^{32}$ erg and is
followed by a rapid $\sim$ 50\% decrease. Differences between the free
energy and the relative helicity reflect the different physics of energy and
helicity removal: free energy is released (dissipated) in the course of any
(eruptive or confined) flare, whereas helicity is bodily removed in the
course of eruptions only (excluding smaller amounts that escape by the
unwinding of ``open" field lines in case of interchange reconnection with
helical ``closed" lines). Indeed, there are occasional decreases of free
magnetic energy that correspond to the energy contents of associated
non-eruptive flares when helicity does not show any significant changes. A
much more significant increase in free magnetic energy than helicity also
occurs, immediately after the X-class flare, with an enhanced energy flux
during that period, as also Figure~\ref{hmdave} indicates. After its peak,
free energy decreases with a faster ($\sim$ 50\%) rate than helicity, which
only drops by 30\%. The initial stages of this considerable decrease, as
discussed in Section~\ref{hel-evol}, are due to fragmentation and
large-scale magnetic field re-organization while later this decrease is the
result of intense, but non-eruptive, flaring activity, as
Figure~\ref{ecevol} shows. Small-amplitude enhancements of free energy
during this last day of observations coincide with similar enhancements of
the energy injection rate.

Following the studied 5-day observing interval, there are still
sufficient budgets of both free magnetic energy and relative helicity left
in NOAA AR 11158 to further power flares and eruptions. Indeed, as the {\em
GOES} event catalogue indicates, two more M-class flares (M6.6 and M1.0) and
31 C-class flares occurred before the AR finally crossed the western solar
limb, concluding its first disk passage. When the AR reappeared in the
eastern limb (as NOAA AR 11171) it was a decaying AR with no sunspots,
having two well separated, disperse opposite polarity regions and a long
H$\alpha$ filament present along the PIL. The AR concluded its
impressive flaring activity, according to the {\em GOES} event catalogue,
with three more flares (M1.5, C4.5 and C6.4), occurring within a couple of
days after its reappearance.

In summary, our magnetic-energy and helicity calculation can provide a
plausible interpretation of the evolution in NOAA AR 11158, in qualitative
agreement with independent observational and modeling/data-analysis studies.
The AR builds rapidly large budgets of right-handed helicity and free energy
sufficient to power several confined and eruptive flares. The overall
dynamical evolution of the AR is a combination of flux emergence injecting
free energy and helicity and flares/eruptions that dissipate or remove parts
of these budgets. In some cases we also witness ``global'' (i.e., AR-wide)
re-organizations of the region's magnetic configuration. This usually
follows swarms of flaring activity that affects free-energy and
relative-helicity budgets. However, reorganizations occur also in the
absence of flares and this may be due to fragmentation, in case flux
emergence ceases, or possibly due to large-scale readjustments of the AR
within the global solar magnetic field.

\subsection{Energy -- helicity diagram of NOAA AR 11158}
\label{EHdiagram}

Recently, \cite{tzio12} introduced the ``energy-helicity'' (EH) diagram of
solar ARs, demonstrating a monotonic correlation between the NLFF free
magnetic energy and relative magnetic helicity, as well as respective
thresholds of $\sim 4 \times 10^{31}$ erg and $\sim 2 \times 10^{42}$ Mx$^2$
for an AR to host major (M-class and higher), generally eruptive, flares. A
plausible question is whether this diagram can be reproduced by NOAA AR
11158 alone.

\begin{figure}
\includegraphics[width=\linewidth]{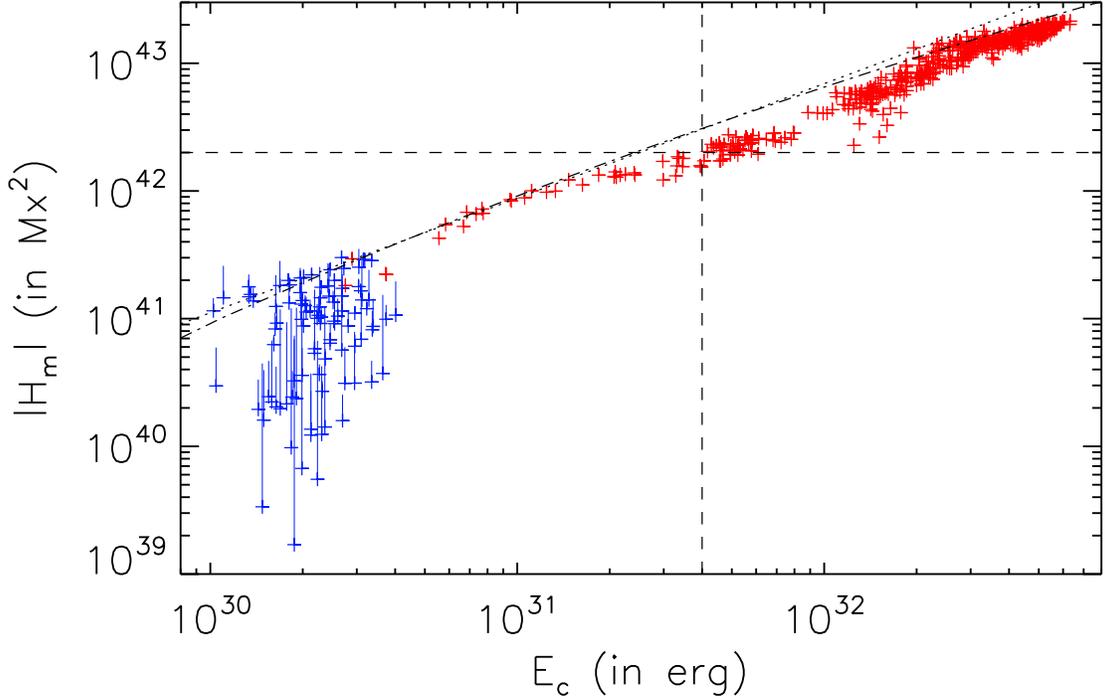}
\caption{The energy-helicity diagram for NOAA AR 11158, comprising of the
AR's free magnetic energy and the respective magnitude of the relative
magnetic helicity. Blue crosses indicate energy and helicity values for the
first 20 hours of 2011 February 12 (see text), while error bars for the
absolute relative helicity (only to higher values) are indicated with blue
lines. Dashed lines indicate the previously estimated thresholds for
relative magnetic helicity ($\sim 2 \times 10^{42}$ Mx$^2$) and free
magnetic energy ($\sim 4 \times 10^{31}$ erg) above which ARs host major
flares \citep{tzio12}. The dotted and dash-dotted lines denote the derived
least-squares best fit and the least-squares best logarithmic fit of
\citet{tzio12} (Equations~(3) and (4), respectively, of that work).}
\label{enhel}
\end{figure}

Figure~\ref{enhel} shows the compiled EH diagram for NOAA AR 11158.
Magnetograms corresponding to the first 20 hours of February 12 (blue
crosses), when the AR has not yet built considerable budgets of magnetic
helicity and free energy (Figures \ref{hmevol}, \ref{ecevol}), show low,
scattered values for both quantities. For this period, there also seems to
exist an exponential increase of  relative helicity with respect to free
energy. If this behavior is real, considering of course the significant
uncertainties, it would indicate that ARs tend to accumulate magnetic
helicity at higher rates than free magnetic energy. However, this behavior
might also be an artifact of our NLFF method that gives rise to exactly zero
free energies for exactly potential fields; this is not completely
guaranteed for helicity, particularly in case of existing flux imbalances
\citep{geo12a}. Concluding, it is rather difficult to disentangle between
these two different possibilities.

Once the active region has accumulated significant helicity and energy
budgets (red crosses) we find that the relative helicity and free magnetic
energy budgets follow quite well the derived least-squares best fit and the
least-squares best logarithmic fit \citep[Equations~(3) and (4),
respectively, of][]{tzio12}. This nearly monotonic dependence holds for a
large range of helicities and energies. Furthermore, as Figures \ref{hmevol}
and \ref{ecevol} suggest, no major flare occurs before {\em both} the
helicity and the energy thresholds are crossed. In fact, not even C-class
flares occur in the AR before these thresholds are exceeded. Once exceeded,
however, a major (M-class) flare occurs within hours.

The single-AR EH diagram of Figure \ref{enhel}, relying solely on
magnetograms from NOAA AR 11158, strongly attests to the validity of the EH
diagram of \citet{tzio12}. It further adds support to the conclusion that
{\em both} magnetic free energy and helicity are important for ARs,
particularly flaring ones.

\subsection{Amplitudes and timing of mutual and self helicity/energy terms}
\label{self}

As shown in Figures \ref{hmevol}, \ref{ecevol} and discussed in Section
\ref{ehbudget}, the mutual terms of the relative helicity and free energy in
NOAA AR 11158 are much larger than the respective self terms. This is not a
new result; \cite{reg05} found from NLFF field extrapolations that the
mutual helicity in NOAA AR 8210 accounts for $\sim 95$\% of the total
helicity budget. In NOAA AR 11158, mutual helicity accounts for even more of
this budget: mutual-helicity terms are typically $\sim 10^3$ times larger
than self-helicity terms, thus accounting for $\sim 99.9$\% of the total
helicity budget. Mutual terms in the free magnetic energy typically account
for even more ($\sim 99.93$\%) of the respective budget. We find, therefore,
that mutual terms caused by the interaction of different NLFF flux tubes
overwhelmingly dominate the twist and writhe contributions of these flux
tubes in the budgets of magnetic helicity and energy. Let us clarify at
this point that the true values of the mutual-to-self helicity and energy
ratios can be determined only in case of fully resolved magnetic
configurations. For $N$ discrete flux tubes (Equations (\ref{Ec_fin}),
(\ref{Hm_fin})) we have $N$ self- and $N^2$-$N$ mutual terms. Changing $N$
due to increasing or decreasing, but always imperfect, spatial resolution,
will modify the number of terms and hence the values of the ratios. In the
extreme case of $N=1$, meaning that the observing instrument resolves a
single flux tube, there will be no mutual terms. The above point taken, it
appears that for $N>1$ or $N\gg1$ independent works invariably find a clear
dominance of mutual over self terms. This should be a real effect.

\begin{figure}
\centerline{\includegraphics[width=0.7\linewidth]{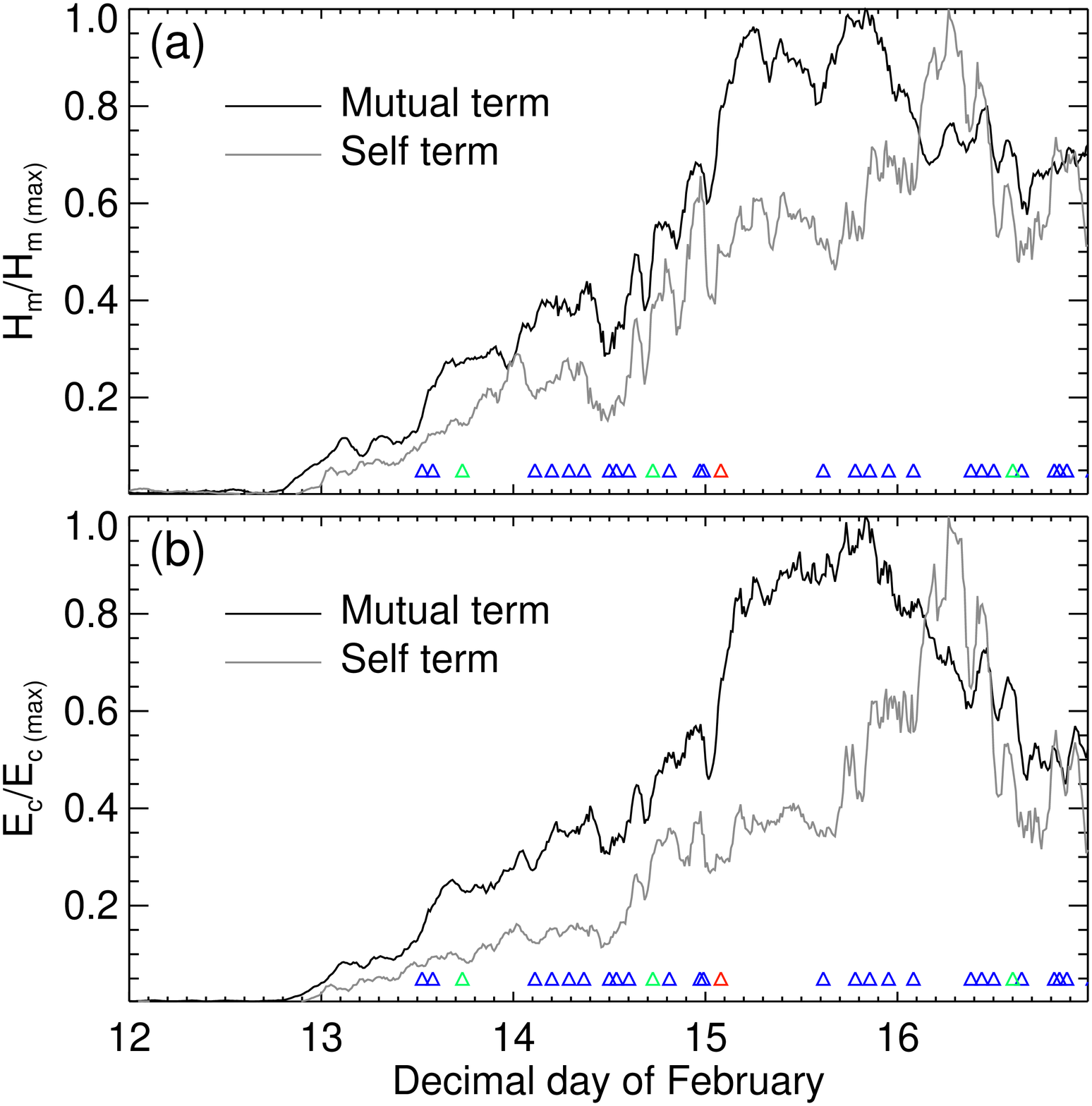}}
\caption{Timeseries of the normalized -- with respect to its maximum --
mutual (black) and self (grey) terms of the relative magnetic helicity (a)
and free magnetic energy (b) in NOAA AR 11158. Colored triangles denote the
onset times of flares in the AR, with blue, green, and red denoting C-, M-,
and X-class flares, respectively.}\label{ms_timing}
\end{figure}

Another interesting result occurs when one studies the {\it relative timing}
between mutual and self terms of helicity and energy. This is exemplified in
Figure~\ref{ms_timing}, where we have normalized mutual and self terms of
helicity and energy by their respective maxima. We find, then, that (i)
there is a {\it hysteresis} in the buildup of self terms of helicity and
energy with respect to these quantities' mutual terms, and (ii) the buildup
rate of self terms of helicity and energy is generally {\it lower} than that
of the respective mutual terms, with the self-term of the free energy
building up at a slightly lower rate than the self term of the relative
helicity. Rough, qualitative features of the mutual-term timeseries are
reproduced by self-term timeseries for both quantities, but at a delay
ranging between several to $\sim24$ hours. We also find remarkable
coincidence in the times of peaks of mutual (helicity and energy) terms, as
well as in the respective peak times of self (helicity and energy) terms.
Mutual terms peak $\sim12$ hours earlier than self terms.

To our knowledge, this is the first time that this result, obviously enabled
by the unparalleled cadence of the SDO/HMI magnetogram data, is reported. We
consider this an important finding because it attests to a possible {\it
conversion} of mutual to self helicity and energy. More specifically, mutual
interactions between an ensemble of flux tubes translate to twist and writhe
of individual flux tubes of the ensemble. We cannot think of any way to
achieve this, other than magnetic reconnection. Indeed, magnetic helicity is
roughly conserved in the course of reconnection forcing interacting
pre-reconnection flux tubes to become more helical after reconnection
because they interact less than before. Naturally, this reconnection will
force interaction changes to the other flux tubes of the ensemble, as well.
If, however, a sequence of reconnection episodes occur in order to relax a
given magnetic configuration, then the trend appears such that increasingly
helical post-reconnection flux tubes are formed. The same essentially
happens with free energy (more twisted structures), as well, but under the
fact that energy always dissipates in the course of reconnection. {bf
These energy losses} may be the reason for the slightly lower buildup rate
of free energy in Figure~\ref{ms_timing}.

The latter result certainly warrants further investigation.
\cite{geo11} suggested that the small-scale flickering of numerous magnetic
reconnection events occurring routinely along active-region PILs
effectively transforms shear-induced mutual helicity into self helicity. The
relation between PIL formation and the origin of shear has been discussed
extensively by \citet{geo12b}. As a result, a new picture of pre-eruption
active-region evolution appears to emerge, with typically eruptive ARs
characterized by intense PILs building increasingly helical magnetic
configurations along these PILs. This argument is in favor of the formation
of {\it pre-eruption flux ropes} in ARs. Very recently, \citet{pats13}
reported direct evidence of pre-eruption flux-rope formation via a confined
flare, that is obviously a magnetic reconnection episode.

\subsection{Helicity and energy contents of major eruptive flares and the flare-CME connection}
\label{erupt}

An admittedly untenable pursuit of relevant analyses of the past was to
detect discrete, significant (i.e. beyond uncertainties) changes in free
magnetic energy and relative magnetic helicity that could be related to
eruptions on a case-by-case basis. Obviously this was attempted under the
assumption that eruptions have energy and helicity contents that are
significant fractions of the respective budgets possessed by the host AR. In
this respect, \cite{metc05} were unable to determine a discrete decrease of
the free magnetic energy in NOAA AR 10486 as a result of a X10 flare in the
region, by applying the Virial theorem to Imaging Vector Magnetograph (IVM)
data. Using vector magnetograms from the Spectropolarimeter (SP) of the
Solar Optical Telescope (SOT) onboard Hinode with a cadence of a few hours,
on the other hand, \cite{geo12a} were able to estimate a change in helicity
and energy of the order $5 \times 10^{42}$ $Mx^2$ and $2 \times 10^{32}$
erg, respectively, apparently related to a X3.4 flare triggered in NOAA AR
10930. The calculation was made as a proof of concept for the NLFF helicity
and energy calculation of Section \ref{newmethod}. However, given the crude
cadence of the SOT/SP data, that result was handled with care.

\begin{figure}
\includegraphics[width=\linewidth]{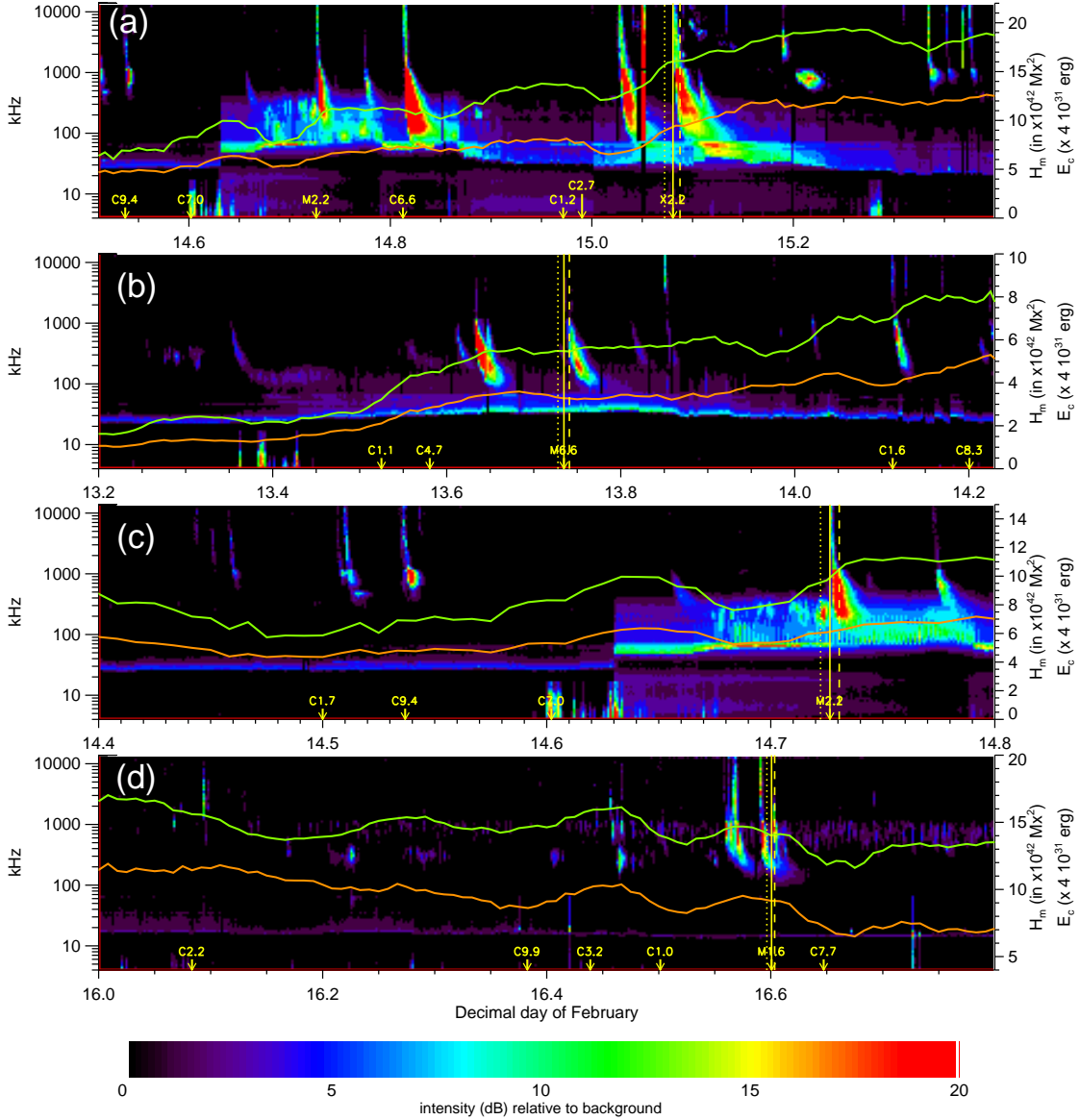}
\caption{72-minute averages of the relative magnetic helicity $H_m$ (green
curves) and free magnetic energy $E_c$ (orange curves) around the times of
the four largest eruptive flares triggered in NOAA AR 11158. These flares
are the (a) X2.2 flare 15-Feb-2011, (b) M6.6 flare of 13-Feb-2011, (c) M2.2
flare of 14-Feb-2011, and (d) M1.6 flare of 16-Feb-2011. Onset times, peaks,
and end times of these flares as registered by GOES are shown by dotted,
solid, and dashed lines, respectively. Peak times of additional flares
triggered in the AR are also indicated. Overplotted for reference are
co-temporal {\em Wind}/WAVES frequency--time radio spectra.} \label{cmes}
\end{figure}

We consider the high-cadence SDO/HMI data of this study ideal for such an
investigation. Using NLFF field extrapolations, \cite{sun12a} detected a
decrease of $\sim 3.4 \times 10^{31}$ erg after the X2.2 flare of 2011
February 15 in NOAA AR 11158. In this study we attempt to detect changes in
both free magnetic energy and relative magnetic helicity. Moreover, we do
not restrict the analysis to the X-class flare alone: instead, we study the
helicity and free-energy evolution in the course of the four (4) largest
eruptive flares in NOAA AR 11158 during the 5-day observing interval. The
results are shown in Figure \ref{cmes} and are overplotted on co-temporal
{\it Wind}/WAVES frequency-time spectra for reference. All shown curves are
72-minute averages. Given the small uncertainties in helicity and free
energy (Figures \ref{hmevol}, \ref{ecevol}), all notable changes that last
much longer than the HMI cadence of 12 minutes should be considered
significant and stemming purely from changes in the calculated connectivity
matrices.

Figure \ref{cmes} leads us to three major conclusions: first, in at least
three of the four eruptive flares studied (Figure \ref{cmes}a, c, and d)
there is a significant decrease in the magnitudes of the AR's relative
helicity and free energy. Second, and perhaps more importantly, these
decreases start occurring {\it well before} the onset of the flare and the
projected launch time of the associated CME. Third, decreases are not
instantaneous but {\em transient}, lasting for 2 -- 3 hours. Flares and CMEs
occur toward the end (15 -- 30 minutes before the end) of these dips. After
the end of the dips, the relative-helicity and free-energy budgets follow
the underlying, general trends of these budgets for the period of interest.

The properties of the relative-helicity and free-energy decreases, including
properties of the four eruptive flares, are summarized in Table \ref{tb1}.
The peak amplitude of the helicity and energy decreases in Figure~\ref{cmes}
is viewed as the respective helicity and energy contents of the associated
eruption. In this respect, the eruption related to the X2.2 flare (Figure
\ref{cmes}a) has an energy content of $\sim 8.4 \times 10^{31}$ erg, a
factor of $<$3 larger than the estimate of \cite{sun12a}. The helicity
budget of the respective CME is $\sim 2.6 \times 10^{42}$ $Mx^2$, in
excellent agreement with estimated typical helicity contents of CMEs
\citep{devo00, geo09}. The other three eruptions are associated with energy
releases of the order $10^{31}$ erg to several times this budget. For the
eruptions related to the M2.2 and M1.6 flares (Figures \ref{cmes}c,
\ref{cmes}d, respectively) the helicity content of the CMEs stays close to 2
-- 3 $\times 10^{42}$ $Mx^2$. For the eruption associated to the M6.6 flare
(Figure \ref{cmes}b), however, it is an order-of-magnitude smaller $\sim 3
\times 10^{41}$ $Mx^2$. We are unclear on why this CME is so much weaker but
this is corroborated by the facts that (i) the CME shock appears rather weak
in {\it Wind}/WAVES spectra (at least weaker than the others in Figures
\ref{cmes}a, c, and d), and (ii) the LASCO CME catalog has registered this
CME as ``Poor'' and ``Partial Halo''.

\begin{table}
\begin{tabular}{cccccccccc}
\hline \hline
   & & \multicolumn{3}{c}{Decrease Properties} & & \multicolumn{2}{c}{Flare Properties} &  & \\
\cline{3-5} \cline{7-8}
No & Date & Start & Peak & End & & Onset & Class   & $\Delta H_m$ ($Mx^2$) & $\Delta E_c$ (erg)\\
\hline
1  & 15-Feb-2011 & 23:22$^\dag$ & 00:13 & 02:00 & & 01:44 & X2.2    & $2.6 \times 10^{42}$  & $8.4 \times 10^{31}$ \\
2 & 13-Feb-2011 & 16:25 & 16:50 & 18:10 & & 17:28  & M6.6 & $2.9 \times 10^{41}$ & $1 \times 10^{31}$\\
3 & 14-Feb-2011 & 15:35 & 16:25 & 17:50 & & 17:20 & M2.2 & $2.8 \times 10^{42}$ & $4.9 \times 10^{31}$\\
4 & 16-Feb-2011 & 11:13 & 12:35 & 13:44 & & 14:19 & M1.6 & $2 \times 10^{42}$ & $6.5 \times 10^{31}$\\
\hline \hline
\end{tabular}
\caption{Temporal properties of the calculated decrease in relative magnetic
helicity and free magnetic energy in NOAA AR 11158 for the four largest
eruptive flares that occurred in the AR during the observing interval. The
estimated eruption-related peak decreases $\Delta H_m$ and $\Delta E_c$ are
also provided. All times are universal times. Helicity and energy values
have been obtained by the 72-minute average timeseries of the respective
quantities (Figures \ref{hmevol}, \ref{ecevol}, respectively).\newline
$^\dag$On 14-Feb-2011} \label{tb1}
\end{table}

\begin{figure}
\includegraphics[width=\linewidth]{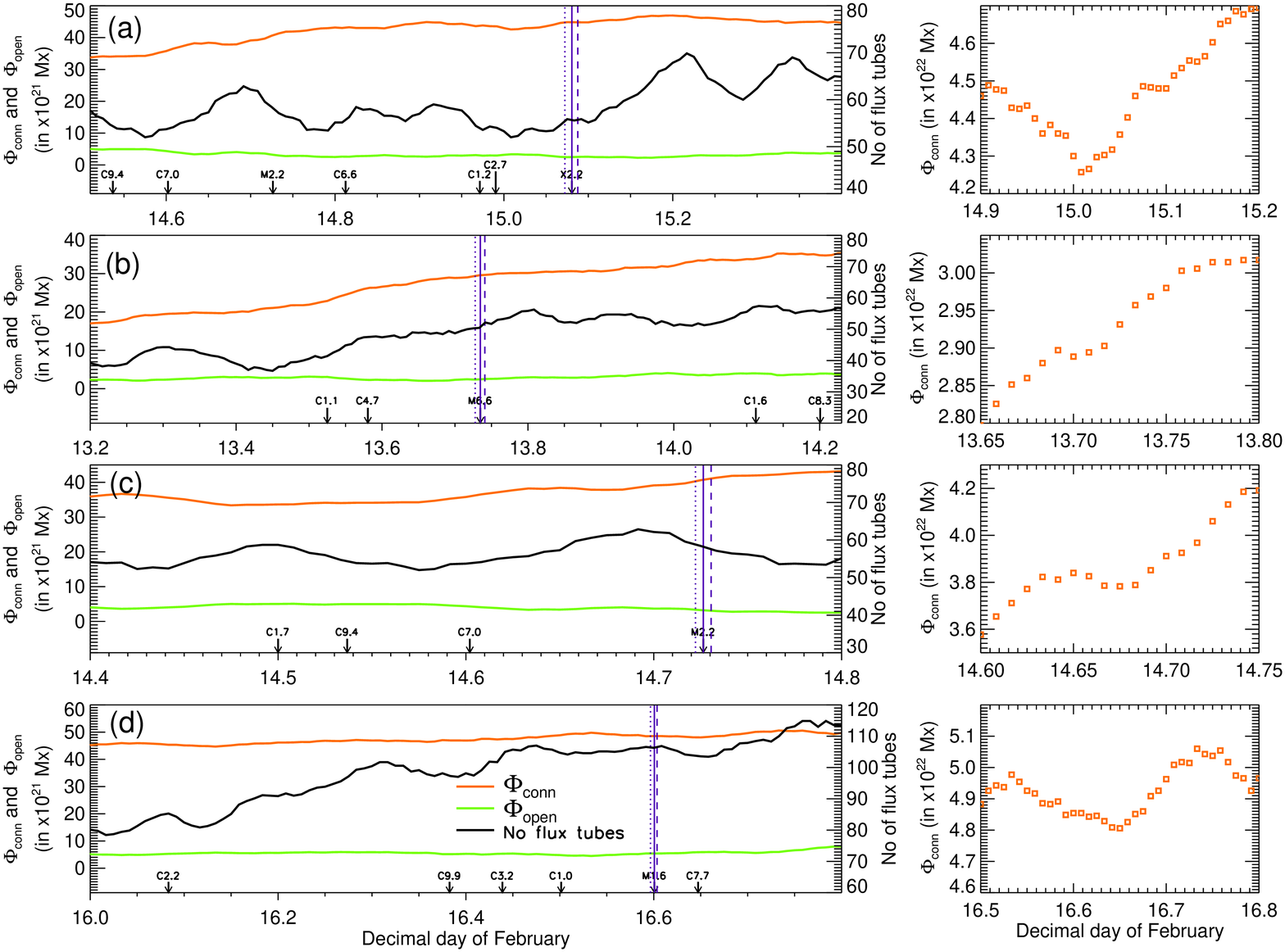}
\caption{Same as Figure \ref{cmes} but showing the 72-minute-averaged
timeseries of the total connected flux $\Phi_{\rm conn}$ (orange curves),
the total open flux $\Phi_{\rm open}$ (green curves), and the number of flux
tubes participating in the connectivity matrix of NOAA AR 11158. For each
plot, the timeseries of the connected flux $\Phi_{\rm conn}$ at the time of
the decrease are shown in the insets at the right.} \label{cmes_2}
\end{figure}

In an attempt to explain why we observe these decreases in the AR's relative
helicity and free energy prior to major eruptive flares, we construct
in Figure~\ref{cmes_2} the respective timeseries for the total connected
flux $\Phi_{\rm conn}$ (i.e., the flux participating in the connectivity
matrix), the total ``open'' flux $\Phi_{\rm open}$ (i.e., the flux closing
beyond the bounds of the AR), and the number of flux tubes participating in
the connectivity matrix. Figure~\ref{cmes_2} shows that co-temporal
with helicity/energy decreases are {\it slight decreases} in the connected
flux $\Phi_{\rm conn}$ for all four eruptive flares. The most significant of
these decreases pertains to the X2.2 flare and associated eruption ($\sim
6$\%; Figures \ref{cmes}a, \ref{cmes_2}a) while the least significant
pertains to the M6.6 flare with the weakest CME ($\sim 0.7$\%; Figures
\ref{cmes}b, \ref{cmes_2}b). These decreases do not seem to be associated
with respective increases in the ``open'' flux $\Phi_{\rm open}$ (except,
perhaps, a very slight increase in Figure \ref{cmes_2}c); this would mean
that previously ``closed'' connections, i.e., connections with both ends
within the AR, now close beyond it. Instead, these decreases imply that the
normal field component $B_z$ that determines the connected flux becomes
somewhat {\em weaker} prior to the eruptions. This is in line with multiple
recent works that advocate for \citep{fis12, hud12} or report
\citep{petrie12, sun12a} enhancements in the horizontal magnetic field and
slighter decreases in the normal field component. Those arguments and
findings mainly refer to the photospheric vicinity of the eruption but, as
we can see from Figure \ref{cmes_2}, they are also detectable over the
entire AR. However, our results do not support an irreversible or permanent
decrease of the connected flux; in this respect, our results run counter to
those works advocating irreversibility, at least in terms of a
$B_z$-decrease. Point taken, notice that \citet{sun12a} did not find an
unambiguously irreversible $B_z$-decrease in the case of the X2.2 flare in
NOAA AR 11158.

Thanks to the unprecedented cadence of the SDO/HMI magnetogram data, our
results help clarify an additional crucial point: so far the above
eruption-related flux changes were confirmed by inspecting the pre- and
post-eruption configurations. Here we deduce that flux-decreases occur
mostly {\it prior} to eruptions, with the eruptions themselves occurring
either at the last stages of these decreases or shortly thereafter. This is
sound evidence toward resolving the elusive flare-CME connection: from our
results, the unstable magnetic structure that will evolve to become a CME
starts ascending {\it clearly} before the flare and the a-posteriori
projected CME launch time. To our understanding, this is the only
possibility that can interpret the connected-flux decreases as changes in
the Lorentz force, but also as contraction / collapse of pre-eruption loops
as the CME progenitor increases in size and ascends faster \citep{vour12}.
The flare is a result of the drastic perturbation caused by this ascension.
Notice that this interpretation has appeared also in earlier works
\citep{zd06} but only now has this picture been clearly supported by
independent observational, modeling, and theoretical works.

The present study includes only a limited sample of eruptive flares. In
followup  works we will further test our findings with larger flare samples
and aim to compare results between eruptive and confined flares.

\section{Discussion and conclusions}
\label{conc}

The magnetic free energy and relative magnetic helicity budgets in NOAA AR
11158 have been studied using high-cadence, constant-quality, space-based
vector magnetograms acquired by the HMI magnetograph onboard SDO. Budgets
have been derived using the novel NLFF method introduced by \cite{geo12a}
that requires {\it single} photospheric (or chromospheric) vector
magnetograms. The constructed timeseries of budgets for free energy and
relative helicity comprise of values independent from each other. For
reference, we use energy and helicity budgets obtained by time-integration
of energy and helicity injection rates, respectively, inferred using the
DAVE4VM photospheric flow field. Successive points in the budget timeseries
are related to each-other in this case. The only alternative way to infer
magnetic energy and helicity budgets is by (NLFF, preferably) magnetic-field
extrapolations, relying also on single vector magnetograms and hence
providing independent budgets in the timeseries. However, both
extrapolations and energy/helicity fluxes require additional nontrivial
information in order to yield energy and helicity budgets (see
Introduction). Much of this information, particularly the NLFF-extrapolated
and the velocity fields, are strongly model-dependent{\footnote{Our choice
for the DAVE4VM velocity field relies on studies that this velocity field,
consistent with the MHD induction equation, is one of the most credible
choices for the photosphere \citep{wels07, sche12}.}}. Therefore, if we can
show that our method reliably calculates energy and helicity budgets at a
given AR without requiring information other than this AR's vector
magnetogram, then this or refined future techniques of this kind might
become methods of choice for the task at hand.

Keeping in mind that the rate-integrated helicity and energy budgets do not
share the same physical meaning with our instantaneous budgets, we find a
qualitative correspondence between the two (Section \ref{ehcomp}). Indeed,
the peak of the integrated helicity in the region is $\sim 2.5 \times
10^{43}$ Mx$^2$ while our instantaneous budget shows a peak $\sim 2 \times
10^{43}$ Mx$^2$. The total energy budget from our method, on the other hand,
peaks at $\sim 1.6 \times 10^{33}$ erg, while the peak of the
flux-integrated budget is lower, $\sim 9 \times 10^{32}$ erg. The
morphologies of the two timeseries are different due to their different
physical origins and meaning. This being said, our NLFF field energy and
helicity budgets depend more sensitively on the AR's eruptive and
non-eruptive evolution. We conclude, therefore, that NLFF methods such as
the one implemented here are indeed important for decoding the dynamical
evolution of ARs from high-cadence vector magnetogram data.

Stepping on this central conclusion, our analysis has led to a meaningful
interpretation of the evolution and the eruptive activity of NOAA AR 11158.
Besides free magnetic energy, analysis provides further evidence that
magnetic helicity is, indeed, an important ingredient of the overall
evolution in solar ARs.

\begin{figure*}
\includegraphics[width=\linewidth]{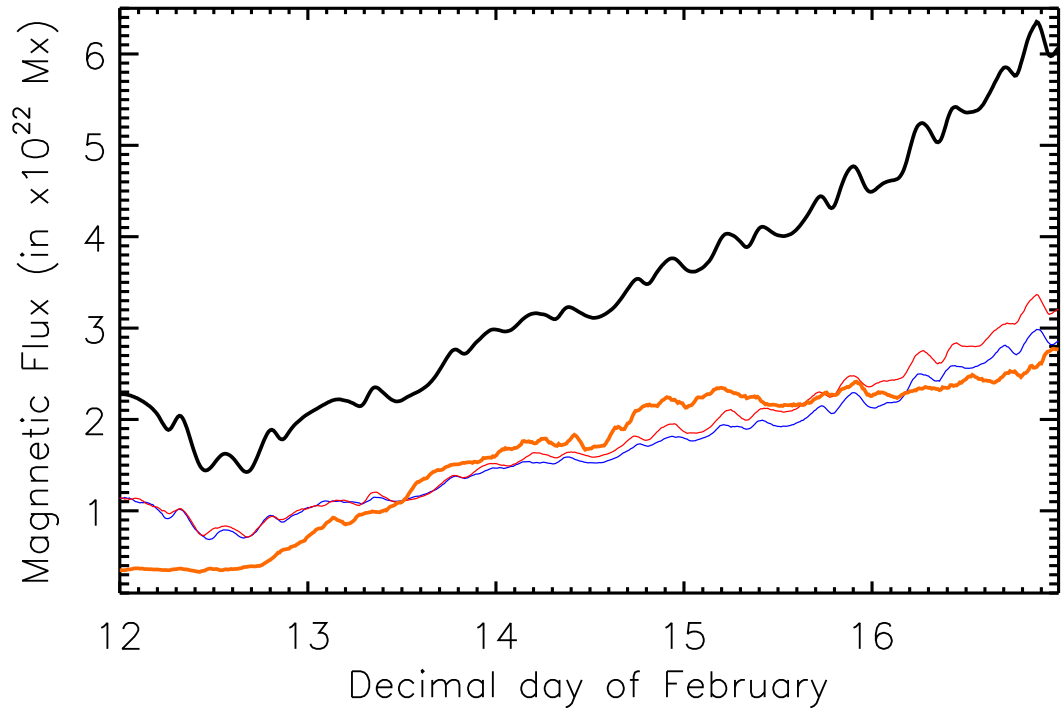}
\caption{Evolution of magnetic flux budgets in NOAA AR 11158. The unsigned
(total flux) is shown by the black curve. The magnitudes of positive- and
negative-polarity fluxes are shown by the red and blue curves, respectively.
The unsigned flux participating in the magnetic connectivity matrix in the
AR (see Section \ref{newmethod} for details) is shown by the orange curve.
For clarity in assessing the long-term evolution of the AR, all shown curves
are 72-minute averages of the actual curves.} \label{flux}
\end{figure*}

The attained budgets of free magnetic energy and relative helicity in NOAA
AR 11158 imply clearly that the AR builds up {\em both} significant amounts
of energy and dominant (in this case positive [right-handed]) helicity
before entering major flaring and eruptive territory. Our method only shows
the increases in the energy/helicity budgets without providing a rigorous
physical investigation of why this happens except, of course, the increasing
size and complexity of the magnetic connectivity matrix that is fuelled by
continuous flux emergence in the AR (Figure \ref{flux}). Previous
observational works that reported shearing motions along the AR's PIL and
rotating sunspots (Section \ref{liter}) complement the picture and our
results. The AR hosts major eruptions when both previously inferred
free-energy and relative-helicity thresholds, $4 \times 10^{31}$ erg and $2
\times 10^{42}$ Mx$^2$, respectively, are exceeded \citep{tzio12}. In
addition, the constructed -- from the 600 magnetograms of this AR alone --
EH diagram clearly follows and further validates the monotonic correlation
between the two quantities found by \cite{tzio12} from the study of 42 ARs
(Section \ref{EHdiagram}).

Superposed on the overall increasing tendency of free-energy and
relative-helicity budgets throughout the 5-day observing interval are
prolonged, significant decreases of these budgets that last for several
hours. We have identified two different reasons for these dips: first, a
``large-scale'', AR-wide re-organization of the magnetic configuration and,
second, confined/eruptive flaring activity in the AR.

\begin{figure}
\includegraphics[width=\linewidth]{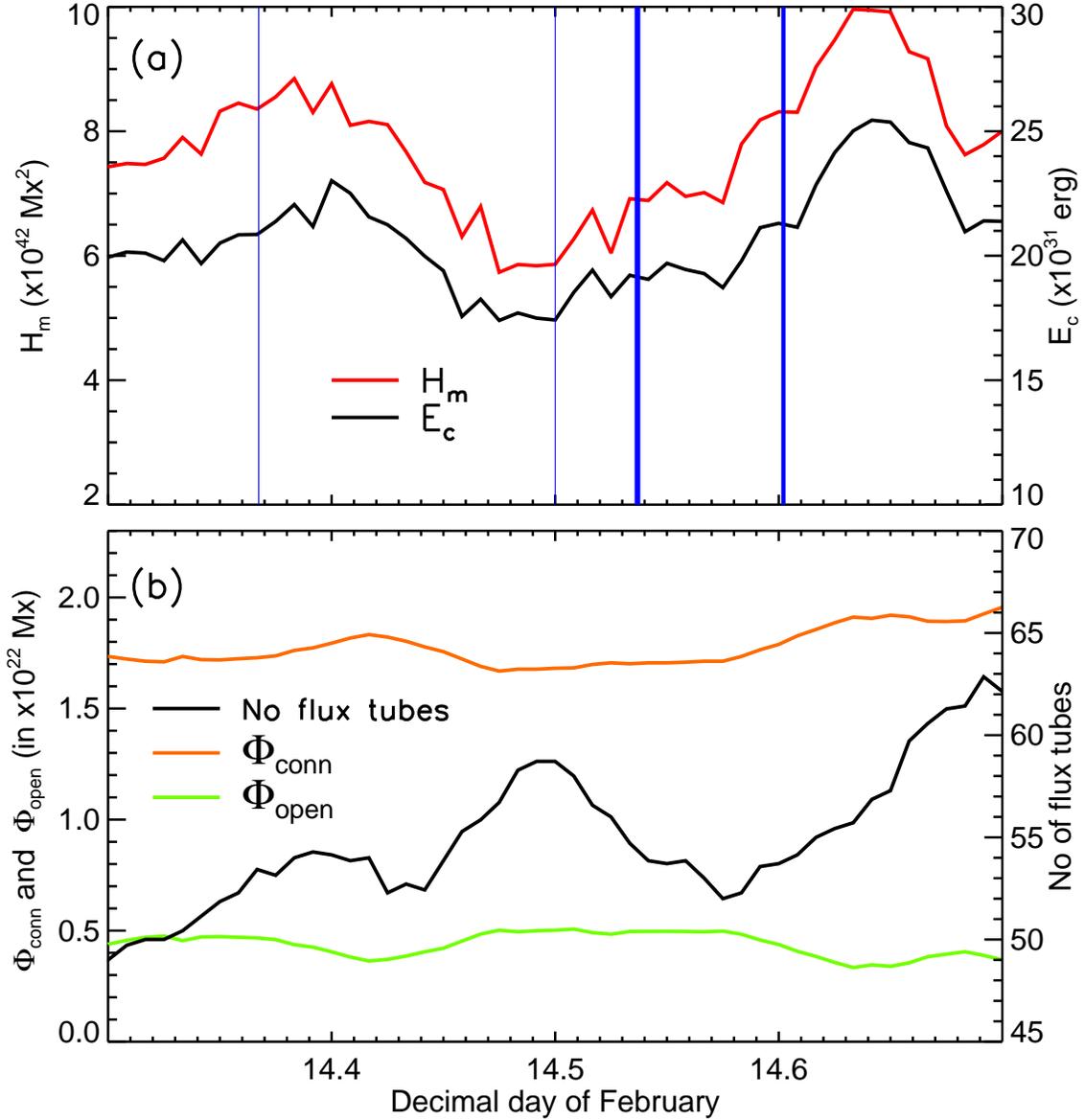}
\caption{72-minute average temporal profiles of (a) relative magnetic
helicity and free magnetic energy, and (b) unsigned connected flux
$\Phi_{\rm conn}$, open flux $\Phi_{\rm open}$ and number of flux tubes
participating in the connectivity matrix around midday on 2011 February 14.
Vertical blue lines in (a) denote the peak times of C-class flares with
their thickness roughly corresponding to the flare magnitude.}
\label{openflx1}
\end{figure}

\begin{figure}
\includegraphics[width=\linewidth]{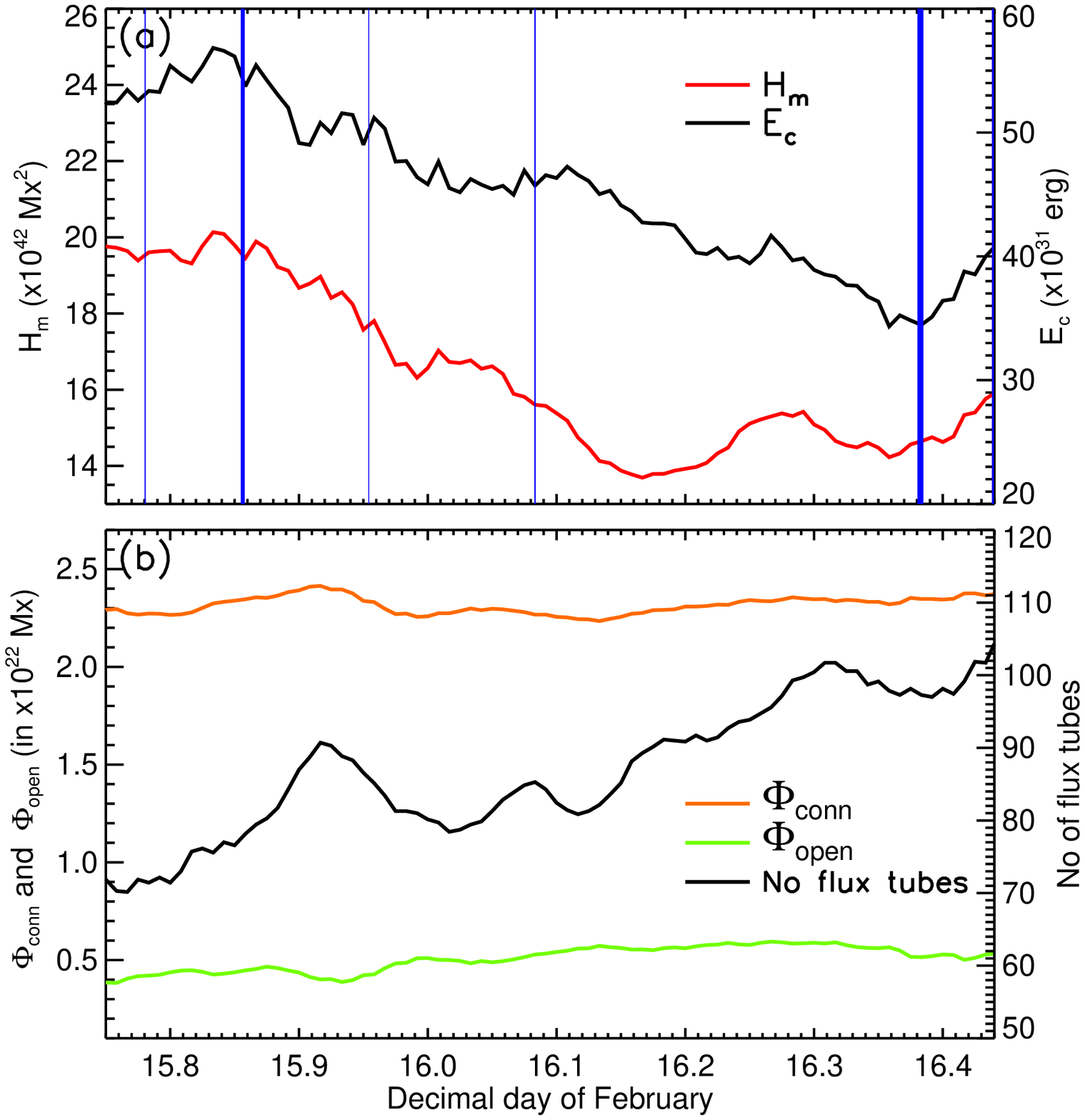}
\caption{72-minute average temporal profiles of (a) relative magnetic
helicity and free magnetic energy, and (b) unsigned connected flux
$\Phi_{\rm conn}$, open flux $\Phi_{\rm open}$ and number of flux tubes
participating in the connectivity matrix between late February-15 and early
February-16. Vertical blue lines in (a) denote the peak times of C-class
flares with their thickness roughly corresponding to the flare magnitude.}
\label{openflx3}
\end{figure}

An AR-wide re-structuring that results in energy/helicity decreases
manifests itself via (i) a decrease of the connected flux $\Phi _{\rm conn}$
participating in the connectivity matrix, accompanied by an increase of the
respective ``open'' magnetic flux $\Phi _{\rm open}$, or (ii) an increase of
the number of active connections (thought as slender flux tubes) accompanied
by a decrease in $\Phi _{\rm conn}$ (Figures \ref{openflx1},
\ref{openflx3}). In the first case, flux closing within the AR up until a
certain time now closes beyond its limits. As ``open'' flux is not currently
included in our energy/helicity calculations, this results in a decrease of
the calculated budgets. In the second case, the AR fragments to more
partitions at smaller $\Phi _{\rm conn}$ leading to the same net result.

An eruption in the AR, on the other hand, will result in a temporarily
smaller $\Phi_{\rm conn}$, apparently due to the eruption-related weakening
of the normal-field component (Section \ref{erupt}). This has been
attributed to the action of the Lorentz force in the eruption area and is
supported by observations. However, the new result in our case is that the
weakening occurs {\it before} the observed onset of the eruption, i.e. the
flare onset and/or the projected CME launch time. We interpret this finding
as solid evidence that the CME progenitor starts ascending well before the
flare. This sheds light on the elusive flare-CME connection and interprets
flares as consequences of the CME-progenitor destabilization, rather than as
triggers of this destabilization. This behavior is clearly seen in three out
of four major eruptive flares the AR hosted within the observing interval.
In the remaining eruptive flare the AR exhibited qualitatively the same
behavior but to a much smaller extent. The most conspicuous decrease of
$\Phi_{\rm conn}$ was $\sim 6$\% and was associated with the X2.2 flare of
2011 February 15, giving rise to a helicity content of the CME of the order
$2.6 \times 10^{42}$ Mx$^2$ and an energy content for the eruption as a
whole of the order $8.4 \times 10^{31}$ erg. If similar results are
confirmed, and we plan to carry out follow-up studies of this type, we
envision a routine calculation of eruption energetics and helicity in the
future. At the moment, nevertheless, it appears safe to conclude that ARs
possess much higher free energy and helicity budgets than those of the
largest eruptions they may ever host. When this is no longer valid, the AR
in question has already entered the final decay phase that will result
in its demise.

Another finding of this study was that eruption-related energy and helicity
decreases are not irreversible or permanent but transient and restored
within several hours by the general trend of the budgets in the AR. The
interpretation of this effect, especially in view of works that show
permanent decreases of $B_z$ and/or increases of the horizontal field,
remains to be investigated.

\begin{figure*}
\includegraphics[width=\linewidth]{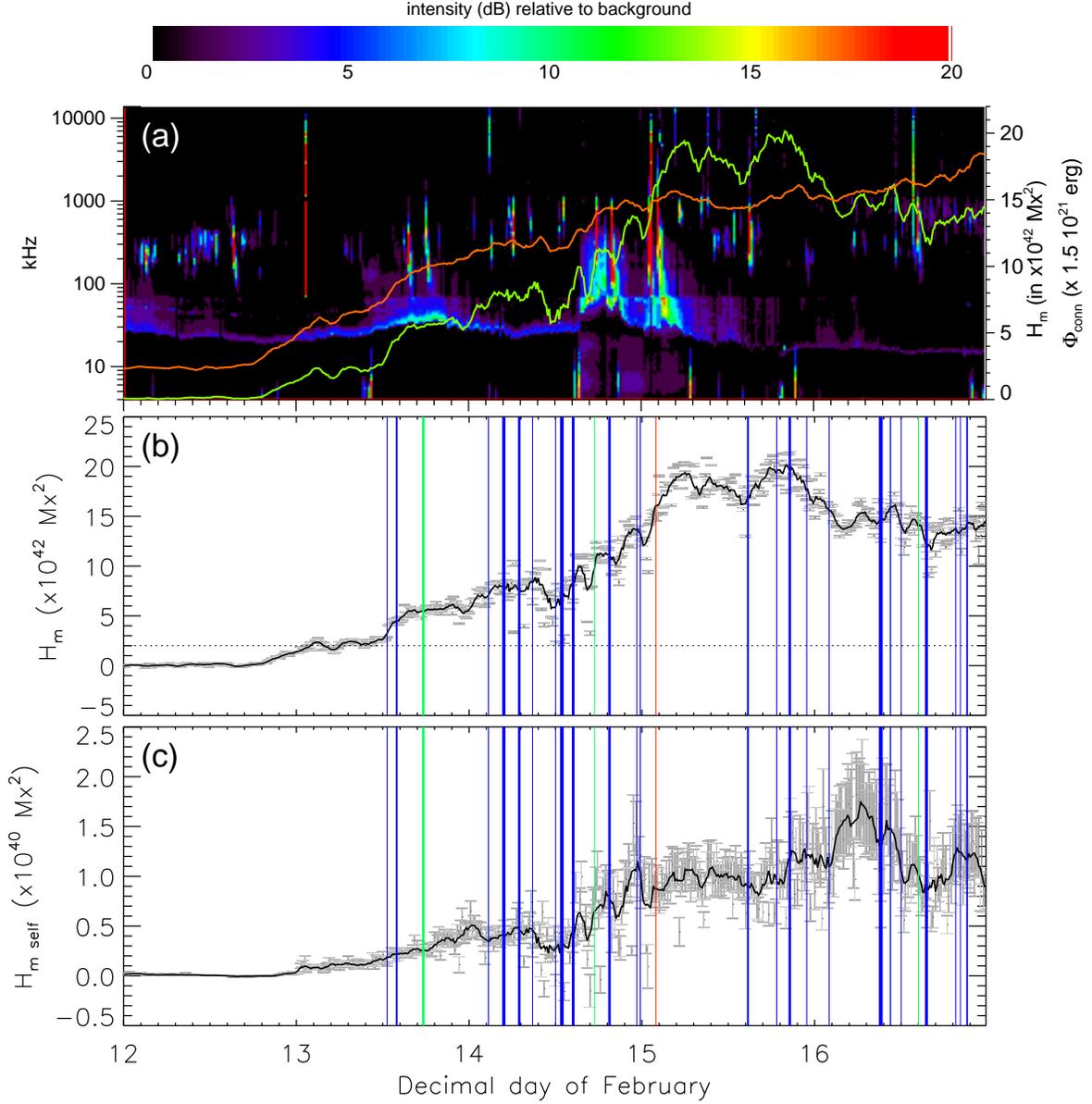}
\caption{(a) 72-minute averages of the resulting NLFF relative magnetic
helicity $H_m$ (green curve) and unsigned connected magnetic flux $\Phi_{\rm
conn}$ (orange curve), overplotted on a frequency-time {\em Wind}/WAVES
radio spectrum for the 5-day observing interval. (b) 72-minute average $H_m$
(solid curve) and actual $H_m$ (grey symbols and error bars) for the same
observing interval. Notice that $H_m$ is indistinguishable from its mutual
term $H_{m_{mut}}$ due to the small magnitude of its self term
$H_{m_{self}}$. (c) 72-minute average $H_{m_{self}}$ (solid curve) and
actual $H_{m_{self}}$ (grey symbols and error bars). Vertical blue, green,
and red lines in (b) and (c) denote the peak times of C-, M-, and X-class
flares, respectively, with their thickness roughly corresponding to the
magnitude of the given flare class. The dotted horizontal line in (b)
indicates the $\sim 2 \times 10^{42}$ Mx$^2$ threshold for relative magnetic
helicity \citep[see text and][]{tzio12}.} \label{hmevol}
\end{figure*}

\begin{figure*}
\includegraphics[width=\linewidth]{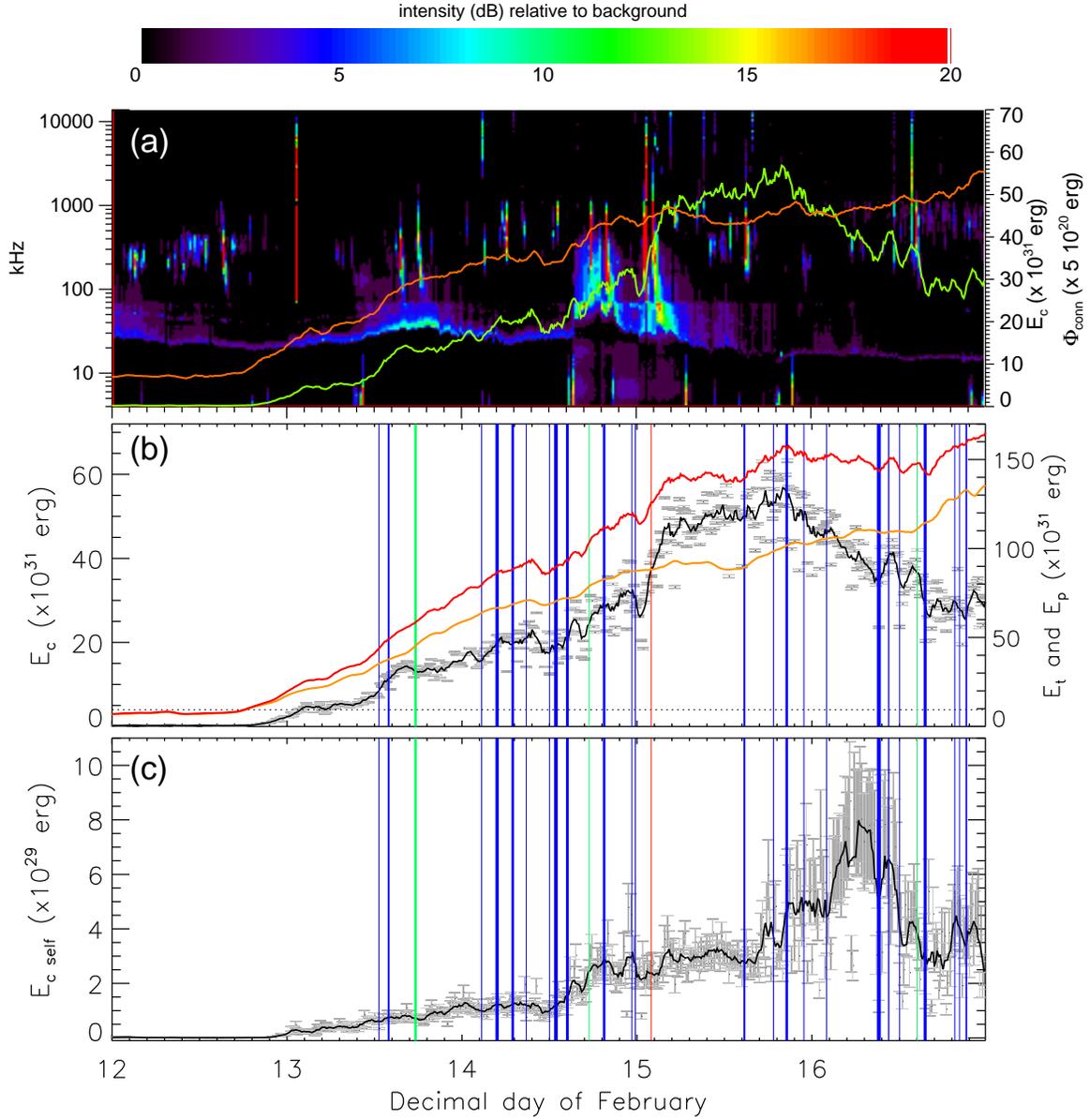}
\caption{(a) 72-minute averages of the resulting NLFF free magnetic energy
$E_c$ (green curve) and unsigned connected magnetic flux $\Phi_{\rm conn}$
(orange curve), overplotted on a frequency-time {\em Wind}/WAVES radio
spectrum for the 5-day observing interval. (b) 72-minute average $E_c$
(solid black curve) and actual $E_c$ (grey symbols and error bars) for the
same observing interval. The reference (potential) and total magnetic
energy, $E_p$ and $E_t$, respectively, are shown by the yellow and red
curves, respectively. Notice that $E_c$ is indistinguishable from its mutual
term $E_{c_{mut}}$ due to the small magnitude of its self term
$E_{c_{self}}$. (c) 72-minute average $E_{c_{self}}$ (solid curve) and
actual $E_{c_{self}}$ (grey symbols and error bars). Vertical blue, green,
and red lines in (b) and (c) denote the peak times of C-, M-, and X-class
flares, respectively, with their thickness roughly corresponding to the
magnitude of the given flare class. The dotted horizontal line in (b)
indicates the $\sim 4 \times 10^{31}$ erg threshold for free magnetic energy
\citep[see text and][]{tzio12}.} \label{ecevol}
\end{figure*}

A further important finding that corroborates the credibility of our
calculations is the AR's evolution during the last day of the observing
interval. Activity in the AR on 2011 February 16 was characterized by
repeated but {\it confined} flaring activity, lacking conspicuous CMEs.
Monitoring the free magnetic energy of the AR (Figure \ref{ecevol}) we
notice a nearly monotonic decrease by $\sim 3 \times 10^{32}$ erg within
less than a day, mostly due to energy dissipation in the course of repeated
flaring. The relative helicity, however, fluctuates around a well-defined
mean value of $\sim 1.5 \times 10^{43}$ Mx$^2$ (Figure \ref{hmevol}) due to
its rough conservation in the course of the confined flares. Such different
behavior stemming from the core properties of magnetic free energy and
relative magnetic helicity accounts for nonlinearities in the EH diagram of
solar ARs.

Concluding, we underline the new, emerging picture of the pre-eruption
evolution in ARs that our results have enabled and previous studies have
envisioned: this picture relies on the timing between mutual- and self-terms
of free energy and relative helicity. We find a significant (much larger
than the observational cadence) {\it hysteresis} of self-term development in
comparison to mutual terms that are developed earlier and at higher buildup
rates (Section \ref{self}). We consider this evidence of a mutual-to-self
{\it conversion} of energy and helicity, rather than an independent,
out-of-phase buildup of both. Moreover, the higher buildup rate of
self-helicity compared to the respective rate of the self-free-energy, in
view also of the rough conservation of helicity in an isolated, non-eruptive
magnetic structure points to a simple (and single) interpretation of this
conversion: {\it magnetic reconnection}. The most profound reconnection area
of an AR is its PIL(s); the stronger the PIL, the more intense the
flickering of magnetic reconnection along it. Reconnection along intense
PILs is due to the action of shear that adds stresses to the already complex
magnetic configuration, while shear is caused by Lorentz force that acts
along the PIL and only there \citep{geo12b}. As hinted by \cite{geo11} and
shown here more convincingly, this logical, hierarchical sequence of
dynamical evolution results in increasingly helical magnetic structures as
PIL evolves. Otherwise put, this may well be independent evidence in favor
of pre-eruption helical flux-rope formation that was hypothesized,
theorized, or observationally reported by multiple previous works. We plan
to further elucidate this result in future follow-up studies.

\acknowledgments We are grateful to P.W. Schuck for valuable, extensive
discussions pertaining to DAVE4VM and its application to this data set. The
data are used courtesy of NASA/SDO and the HMI science team. YL was
supported by NASA Contract NAS5-02139 (HMI) to Stanford University. This
work was supported from the EU's Seventh Framework Programme under grant
agreement n$^o$ PIRG07-GA-2010-268245.

\clearpage

\end{document}